\documentclass[a4paper,fleqn,usenatbib]{mnras}
\usepackage[T1]{fontenc}
\usepackage{ae,aecompl}
\usepackage{subfigure}
\usepackage{color}
\usepackage{graphicx}	
\usepackage{natbib,times}
\usepackage{amsmath}	
\usepackage{amssymb}	
\title[Jiamusi pulsar observations: IV. PSR B0329+54]
      {Jiamusi pulsar observations: IV. The core-weak pattern of PSR B0329+54}

\author[Wang et al.]
       {Tao Wang$^{1,2}$,
         J. L. Han$^{1,2}$\thanks{E-mail: hjl@nao.cas.cn},
         C. Wang$^{1,2}$,
         P. F. Wang$^{1,2}$,
         D. J. Zhou$^{1,2}$
         \\
         $^1$National Astronomical Observatories, Chinese Academy of Sciences,
         Jia-20 Datun Road, ChaoYang District, Beijing 100101, PR China\\
         $^2$School of Astronomy, University of Chinese Academy of Sciences,
         Beijing 100049, PR China  \\
       }

\date{Accepted XXX. Received YYY; in original form ZZZ}

\pubyear{2021}

\begin{document}
\label{firstpage}
\pagerange{\pageref{firstpage}--\pageref{lastpage}}
\maketitle

\begin{abstract}
The bright pulsar PSR B0329+54 was previously known for many years to have two emission modes. Sensitive observations of individual pulses reveal that the central component of pulse profile, which is called core component, is found to be very weakened occasionally for some periods and then recovered. This is the newly identified core-weak mode.
Based on our long observations of PSR B0329+54 by the Jiamusi 66-m telescope at 2250 MHz, we report here that the profile components of individual pulses,  including these for the core and the leading and trailing peaks, are relatedly varying over some periods even before and after the core-weak mode, forming a regular pattern in the phase-vs-time plot for a train of period-folded pulses. The pattern has a similar structure for the core-weak mode with a time scale of 3 to 14 periods. It starts with an intensity brightening at the trailing phase of the core component, and then the core intensity declines to a very low level, as if the core component is drifting out from the normal radiation window within one or two periods. Then the intensity for the trailing components is enhanced, and then the leading component appears at an advanced phase. Such a core-weak mode lasts for several periods. Finally, the core-weak mode ends up with an enhanced intensity at the leading phase for the core component, as if the 
core gradually comes back and finally stays at the phase of the profile center as it 
used to.  

\end{abstract}




\section{Introduction}

Most pulsars have very stable mean pulse profiles. In the core-cones model \citep{ran83}, the central component of mean pulsar profiles is called the core component, 
and the shoulder components are conal components. Pulsar emission may come from random 
sources modulated by a ``window'' \citep{lm+1988,m+1995}, so that the multiple components 
of the mean profile reflect a kind of the averaged ``window'' in neutron star 
magnetosphere \citep{m+1995}. Though the emission geometry of pulsars have been hinted by 
the mean polarization profile, the detailed mechanism and emission processes are not 
known yet. 

In general for bright pulsars, an observed sequence of single pulses shows the strength fluctuations of subpulses and their random or regular locations in the 
longitude phase ranges.  For some pulsars, subpulse emission can be ceased for 
periods, which is called the nulling. Some pulsars have subpulses drifting 
inside the window defined by the mean profile, which is called subpulses drifting. 
Some pulsars have different profiles for some periods, which is called mode-changing.  

PSR B0329+54 is a very bright pulsar with a period of 0.71452s in the northern sky and 
has 5 very distinguished components \citep{lm+1988,ran93}. If single pulses are carefully studied \citep{Gangadhara2001} it shows 9 Gaussian components. These components can be
explained by the cut of sight-line on the central core and the inner cone, the outer cone \citep{ran93,KramerM1994_AA} or very outer cones \citep{Gangadhara2001} in the 
frame of the core and cone model \citep{ran83}. PSR B0329+54 is bright enough so that 
almost every single pulse can be detected by many radio telescopes 
with a very high signal to noise ratio even at 2.3~GHz or 8.6GHz \citep{YanZhen2018_APJ}. 
It is a mode-change pulsar \citep{LyneAG1971_MNRAS,HesseKH1973,Bartel1982,ChengJL2011}, showing two different 
average profiles for the normal mode and the abnormal mode. The duration for 
the normal and abnormal modes occupies about 83\% and 17\% of the whole observation 
time respectively \citep{YanZhen2018_APJ,ChengJL2011}. No subpulses drifting was recognized previously. Very bright subpulses occasionally are detected 
from the core component \citep{YanZhen2018_APJ}. Besides, long observations 
by \cite{whh+20} show no nulling for 22,000 periods. \citet{MitraD2007_MNRAS} noticed the core component occasionally
declines to a very low level in the 325-MHz data  observed by the Giant Metre-wave Radio Telescope, and called it as ``core null''. 
\cite{tyw+2022} detected the low level core emission from single pulses observed at 1.54 GHz by using the NanShan radio telescope, and took it as the core-weak mode.
They discussed the time scales of core-weak mode, and checked possible differences of the low emission mode in the abnormal and normal modes. The influences of the core-weak mode on surrounding single pulses were also
explored. However, details on how the core-weak mode affects the subpulses of individual single pulses were not presented due to the limited signal-to-noise 
ratio (S/N).

\begin{table*}
    \caption[]{Observation sessions for PSR B0329+54 by the Jiamusi 66-m telescope at the S-band.}
    \label{obsDetail}
    \centering
    \begin{tabular}{lccccc}
    \hline
    Obs. Session & Start UTC  & Duration  & Pulses &  Fraction of abnormal & Abnormal pulse range \\
    (yymmdd) & (hh:mm)  & (min.) & (No.) &  ( \% ) & (No. -- No.)\\
    \hline
    20150615 & 19:32 & 31 & 2668  & 0.0 &       --  \\   
    20160221A& 02:29 &300 & 25218 & 6.3 &  23540 -- 25180  \\ 
    20160221B& 14:19 &182 & 15327 & 51.5&      1 -- 5600; 12640 -- 14900 \\
    20160224 & 05:27 &421 & 35400 & 1.4 & 19600 -- 20100 \\ 
    20171108 & 10:41 &794 & 66711 & 17.7&  1 -- 3100; 25600 -- 26080;   27700 -- 34000; 63400 -- 64750   \\
    \hline
    \end{tabular}
\end{table*}

\begin{figure*}
\centering 
\includegraphics[width=0.325\textwidth]{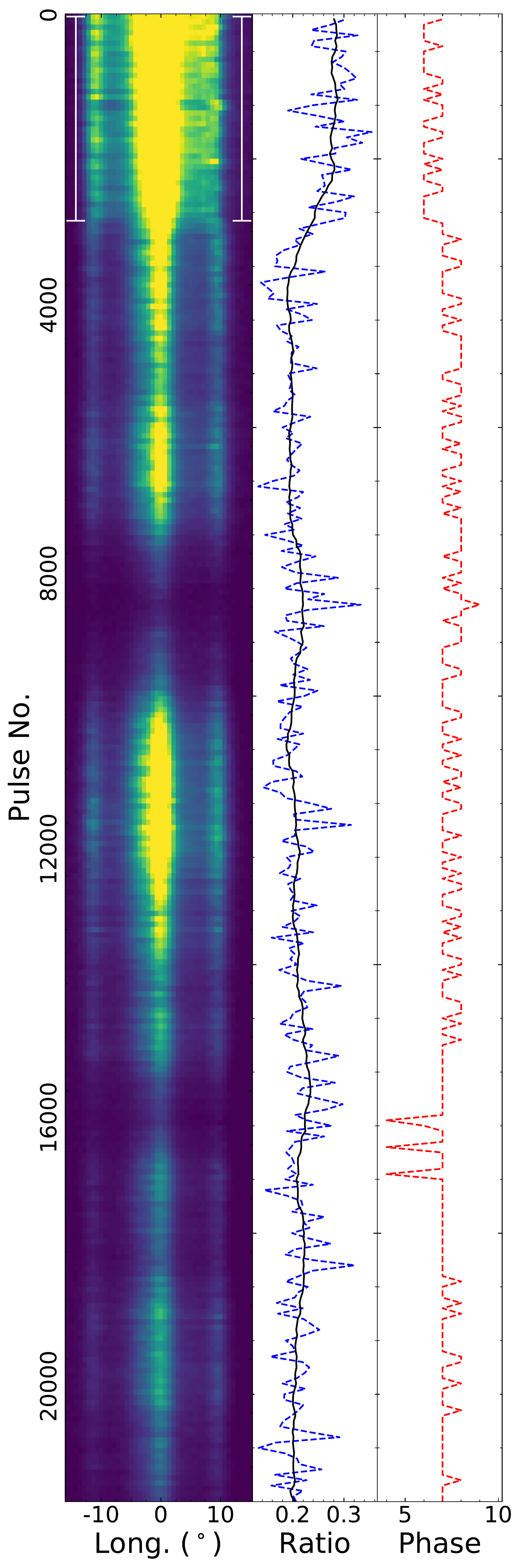} 
\includegraphics[width=0.325\textwidth]{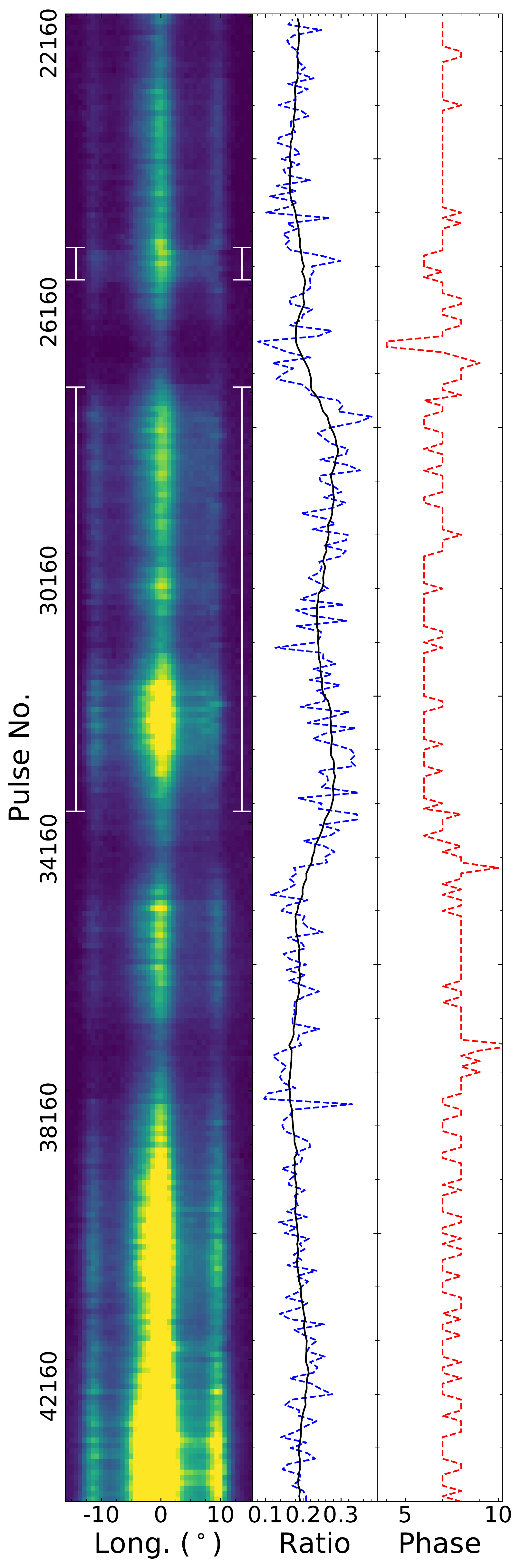}
\includegraphics[width=0.325\textwidth]{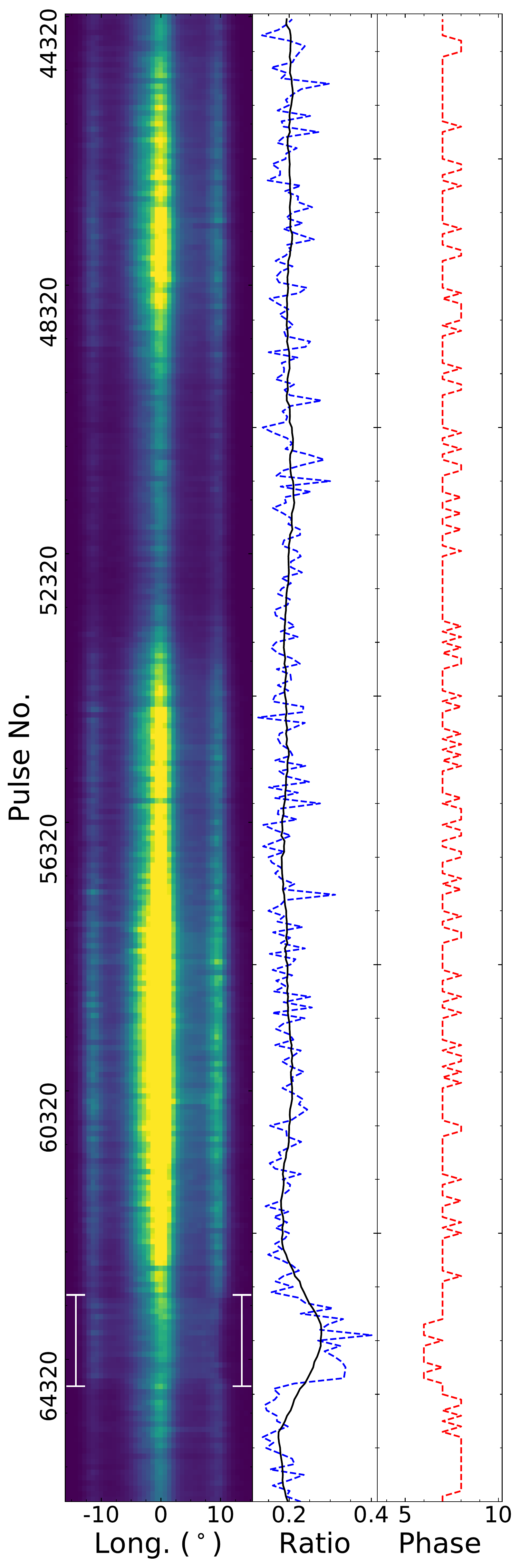} \\
\caption{A long pulse sequence of PSR B0329+54 observed in the 20171108 session
shows several mode-changings. The pulses are integrated for 80 periods in the left panel
to identify the normal and abnormal modes. In the middle 
panel is the variation of the intensity ratio (the dashed line)
between the first peak to the central peak and the black line is the smoothed line. In the right panel is the
phase of the trailing edge at the half of the last peak  in the case of the central peak at the phase of 0$^\circ$. The variations over long timescales is caused by interstellar scintillation \citep{whh+18}.}  
\label{mode-changs}
\end{figure*}

\begin{figure}
\centering 
\includegraphics[width=0.4\textwidth]{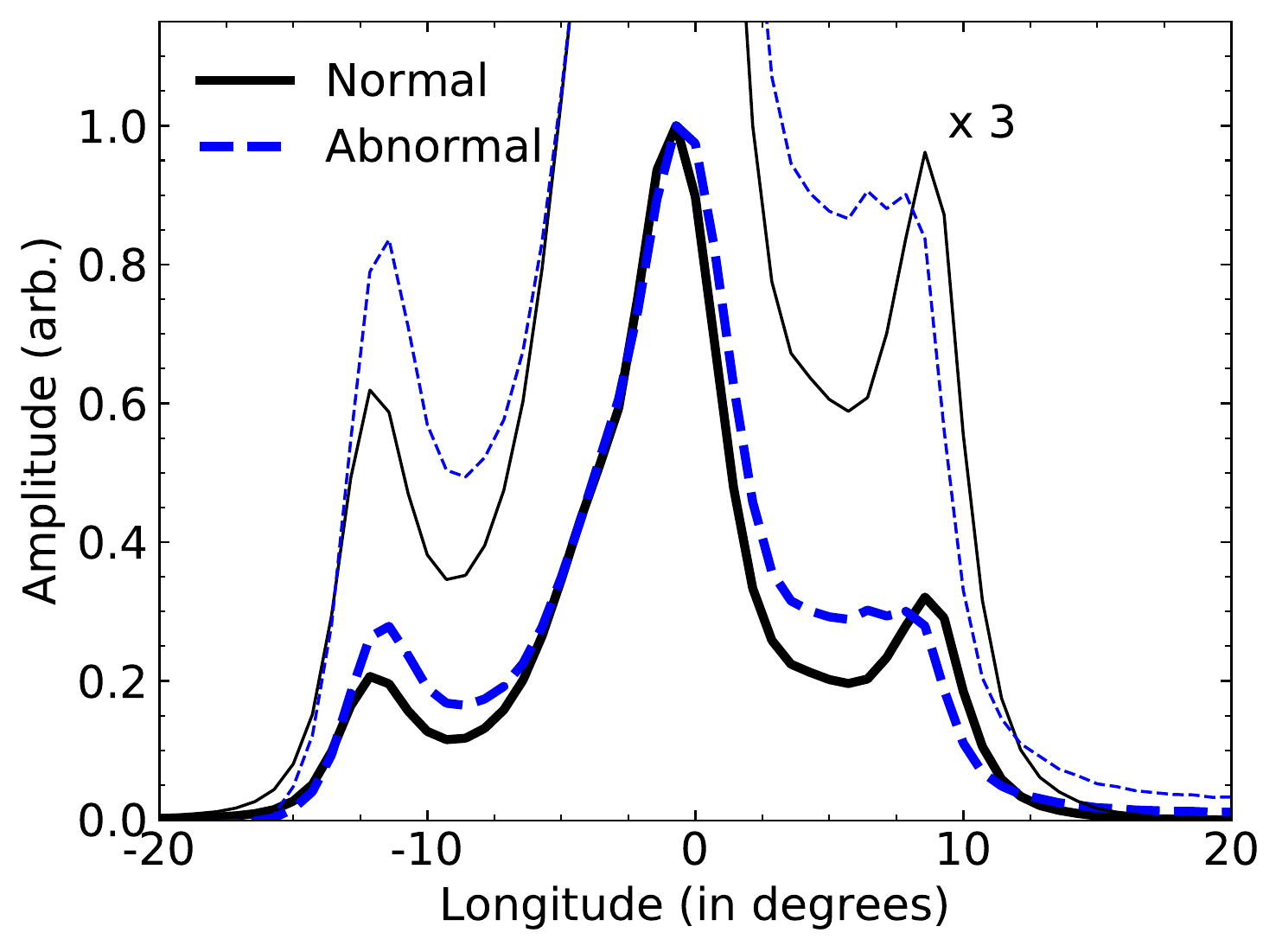}
\caption{Average profiles of normal mode (solid line) and abnormal mode
(dashed line) of PSR B0329+54 obtained in the 20171108 session, normalized by their peak values. The thin lines are magnified versions (3 times).}
\label{aveProf}
\end{figure}

In this paper, we present long sequences of individual pulses of
PSR B0329+54 observed by the Jiamusi 66-m radio telescope at 2.25
GHz. In addition to previously known mode-changing and the newly detected core-weak mode, we recognized the regular pattern in the phase-time plots for individual pulses 
for the core-weak mode, in which the cone and core profile components are relatedly varying for several periods. In Section~\ref{dataArchive}, the observations are briefly introduced. 
In Section~\ref{pattern}, we identify the emission patterns with different time-scales and discuss the properties especially on the intensity variations of the core and conal components. We discuss the probability distribution of patterns for these time-scales. In Section~\ref{DiscussionAndConclusion}, we discuss the physical origin of the core-weak patterns.

\section{Observations and data}
\label{dataArchive}

The Jiamusi 66-m telescope is located at Jiamusi Deep Space Station of
China Xian Satellite Control Center in Heilongjiang province. It has
been used for pulsar observations since 2015 
\citep[see details in][]{HanJun2016}.  The long observations of bright
pulsars in a series of Jiamusi pulsar observation papers have revealed new features for the abnormal emission events  of 
PSR B0919+06  \citep[paper I, by][]{HanJun2016}, scintillation of 10 bright pulsars
\citep[paper II, by][]{whh+18} and nulling of 20 pulsars \citep[paper III, by][]{whh+20}. 

The observations for PSR B0329+54 were carried out by the Jiamusi 66-m
telescope at the S-band with a cryogenically cooled receiver. The signals  are received in dual orthogonal polarisation channels (i.e. the right and left hand circular polarization: R and L) at a center frequency 2253~MHz with a bandwidth of roughly 140~MHz. The digital backend channelizes the down-converted R and L signals into 256 or 128 channels. The power of the RR and LL polarization channels are added, and accumulated and stored in a data file every 0.2~ms or 0.1~ms. The data file is converted to a PSRFITS file later for off-line processing. 

PSR B0329+54 was observed in 5 sessions in the searching mode, see  Table~\ref{obsDetail} for details. The observation sessions on 20150615, 20160221 and 20171108 were carried in night with less radio frequency interference (RFI). We use the function PSRZAP in PSRCHIVE software package \citep{vdo+2012} to clean RFIs. Data of all frequency channels are dedispersed and averaged with weights according to the bandpass to produce the time sequences of data. The ephemeris from the Australia Telescope National Facility Pulsar Catalogue \citep{mhth05} is used to fold the data  so that the pulse phases are well aligned. The longest series of pulse sequences of the 20171108 session is shown in Figure~\ref{mode-changs}. The intensity variations on a time scale of hours are obviously caused by scintillation.



PSR B0329+54 has three prominent peaks in the mean profile (see Figure~\ref{aveProf}) and no conventional nulling was detected previously.
Mode-changing phenomenon of PSR B0329+54 has been detected in the broad frequency 
bands \citep{LyneAG1971_MNRAS,HesseKH1973,Bartel1982,YanZhen2018_APJ}, 
and also in our observations at 2.25 GHz. The average profiles for the normal mode and the abnormal mode are showed in Figure~\ref{aveProf}, which verifies that in the abnormal mode the leading peak and the bridge between the central component and the trailing component becomes stronger and the trailing component shrinks in pulse width. As seen in the longest pulse 
sequence in the 20171108 session in Figure~\ref{mode-changs}, the left panel of this figure shows 
the pulse sequence with an integration of every 80 periods, 
the middle panel shows the ratio of the leading component peak with respect to 
the central component peak (the dashed line) with a smoothing line (the solid line), 
and the right panel shows the variance of phase-width between the 
central component peak and the longitude for the half of the last component peak (the dashed line). 
When the emission is in abnormal mode, the ratio of the leading component peak versus the central component peak becomes larger, and the phase-width between the central peak and the longitude for the half last peak is also reduced, compared to these in the normal mode. The so identified mode-changes are indicated by the white lines in the panel of pulse-stacks. We detected 8 mode-changing events in four sessions, as listed in Table~\ref{obsDetail}. 


\begin{figure}
\centering
\includegraphics[width=0.80\linewidth]{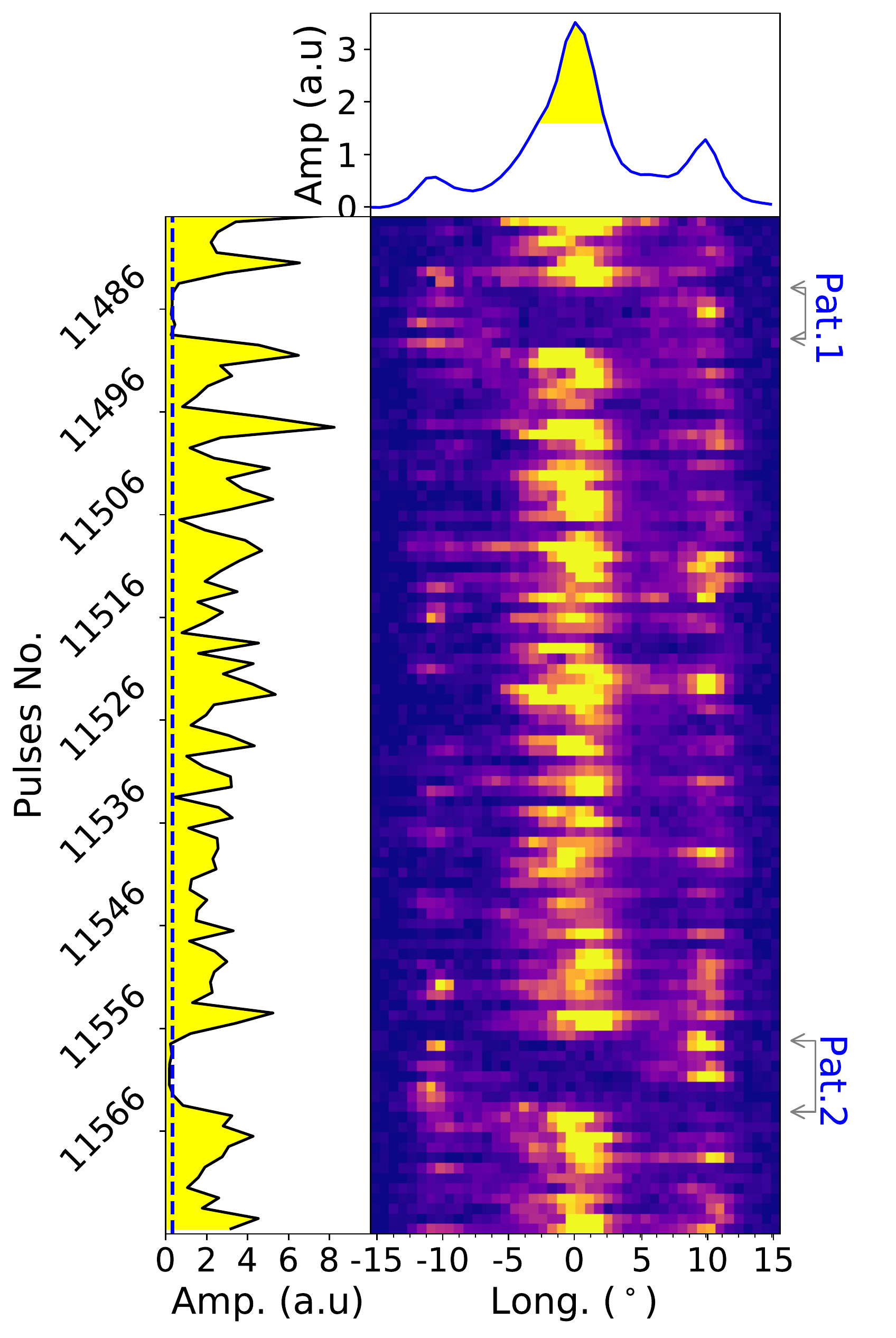} 
\caption{Detailed pulse stacks of PSR B0329+54 observed 
in the 20160204 session showing the two patterns for the core weak mode. The top sub-panel shows the mean profiles (solid line). The core energy variations in the left sub-panel are integrated over the phase bins with the mean pulse greater than the half peak values. The fluctuations of the same number of off-pulse phase-bins are used to derive the root-mean-square for the  uncertainty  ($\sigma$) of core energy variations, and the $3\sigma$ values are indicated by the dashed line in the left sub-panel. 
  }
\label{patternTemplate}
\end{figure}

\begin{table*}
    \caption[]{Number of core-weak emission patterns with different time scales detected in 5 observation sessions.}
    \label{patternNumVSTimescale}
    \begin{tabular}{lcccccccccccc}
    \hline
     Session   &  \multicolumn{12}{c}{Time scales (in period)  } \\
   \hline
    & 3 & 4 & 5 & 6 & 7 & 8 & 9 & 10 & 11 & 12 & 13 & 14 \\
    \hline
    20150616 
    &9 &12 &4 &7 &3 &0 &0 &1 &0 &0 &0 &0\\
    20160221A 
    &105 &66 &70 &44 &33 &17 &8 &3 &1 &1 &0 &1\\
    20160221B 
    &55 &43 &28 &20 &18 &9 &2 &2 &2 &2 &0 &0\\
    20160224  
    &135 &126 &91 &67 &46 &18 &6 &2 &2 &1 &1 &0\\	
    20171108 
    &189 &170 &124 &107 &61 &28 &11 &5 &5 &0 &1 &0\\ \hline
    Total     &493 &417 &317 &245 &161 &72 &27 &13 &10 &2 &1 &1 \\
    \hline
    \end{tabular}
\end{table*}

\section{The core-weak emission pattern }
\label{pattern}

The core-weak mode \citep{tyw+2022} are clearly detected in our single pulse data, two examples showed in Figure~\ref{patternTemplate}. We find that PSR B0329+54 exhibits a peculiar behaviour for the core-weak mode that the core component suddenly becomes very weak at 2.25GHz. More intriguing is that the core components keeps weak for 3 to 14 periods, with clearly well-organized patterns for profile components in the time-phase plot for the pulse stacks, like Pattern No. 2 between No.11555-No.11567 in Figure~\ref{patternTemplate}. We group these patterns according to the number of core-weak periods for the weak core mode, and made the statistics as listed in Table~\ref{patternNumVSTimescale}. All such patterns are recognized by using the amplitude filter method (see Appendix). Finally we get 1759 core-null patterns in total (see Table~\ref{patternNumVSTimescale}). 

\begin{figure}
\centering
\includegraphics[width=0.8\linewidth]{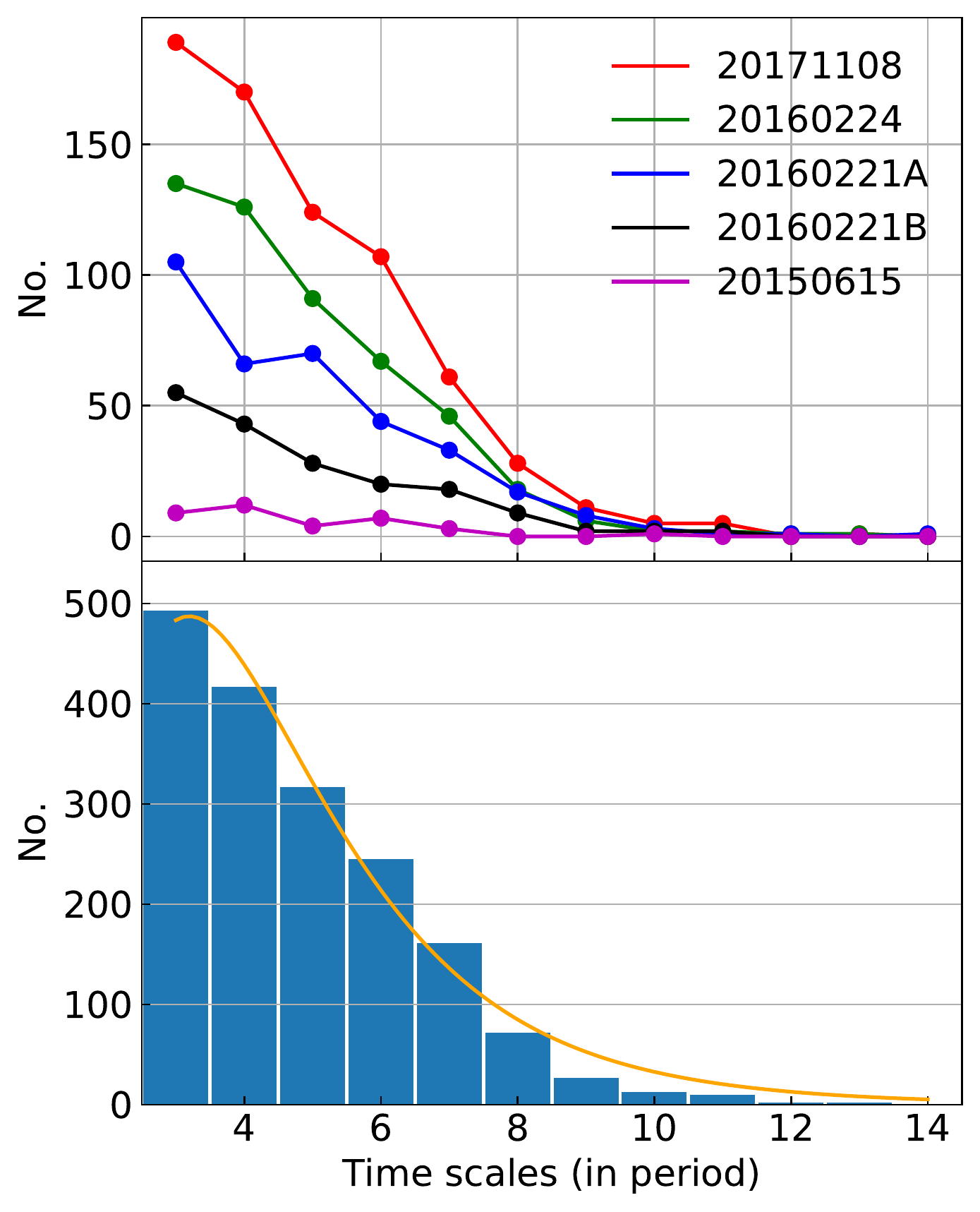} 
\caption{The statistics of core-weak emission mode with different timescales in the 5 observation sessions. The bottom panel shows the total number of all 5 observation sessions against timescale, fitted by a log-normal distribution.}
\label{patternNumDis}
\end{figure}

\begin{figure*}
\centering
\subfigure[timescale=3]{
\label{fig2-a}
\includegraphics[width=0.32\linewidth]{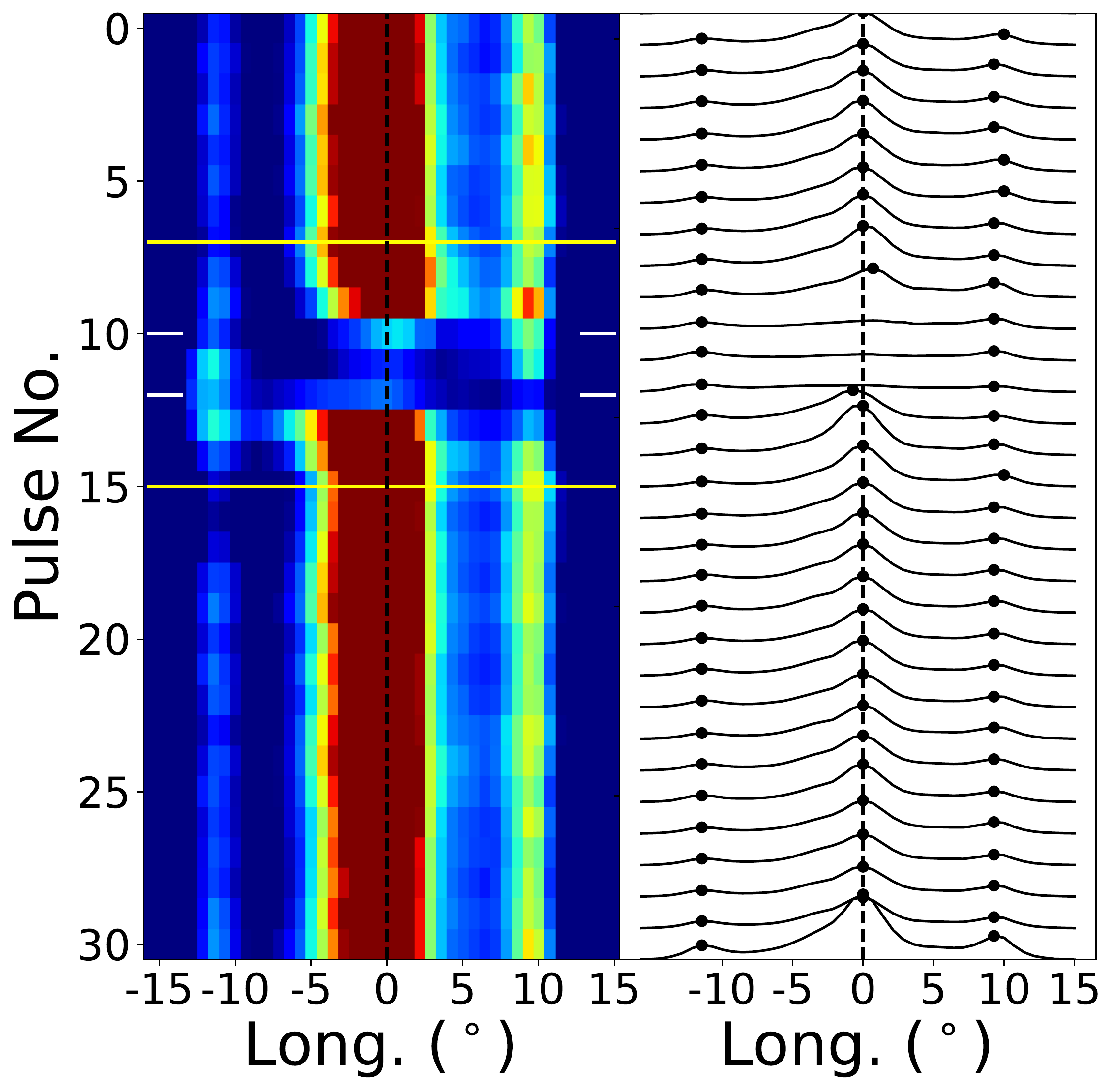}}
\subfigure[timescale=4]{
\label{fig2-b}
\includegraphics[width=0.32\linewidth]{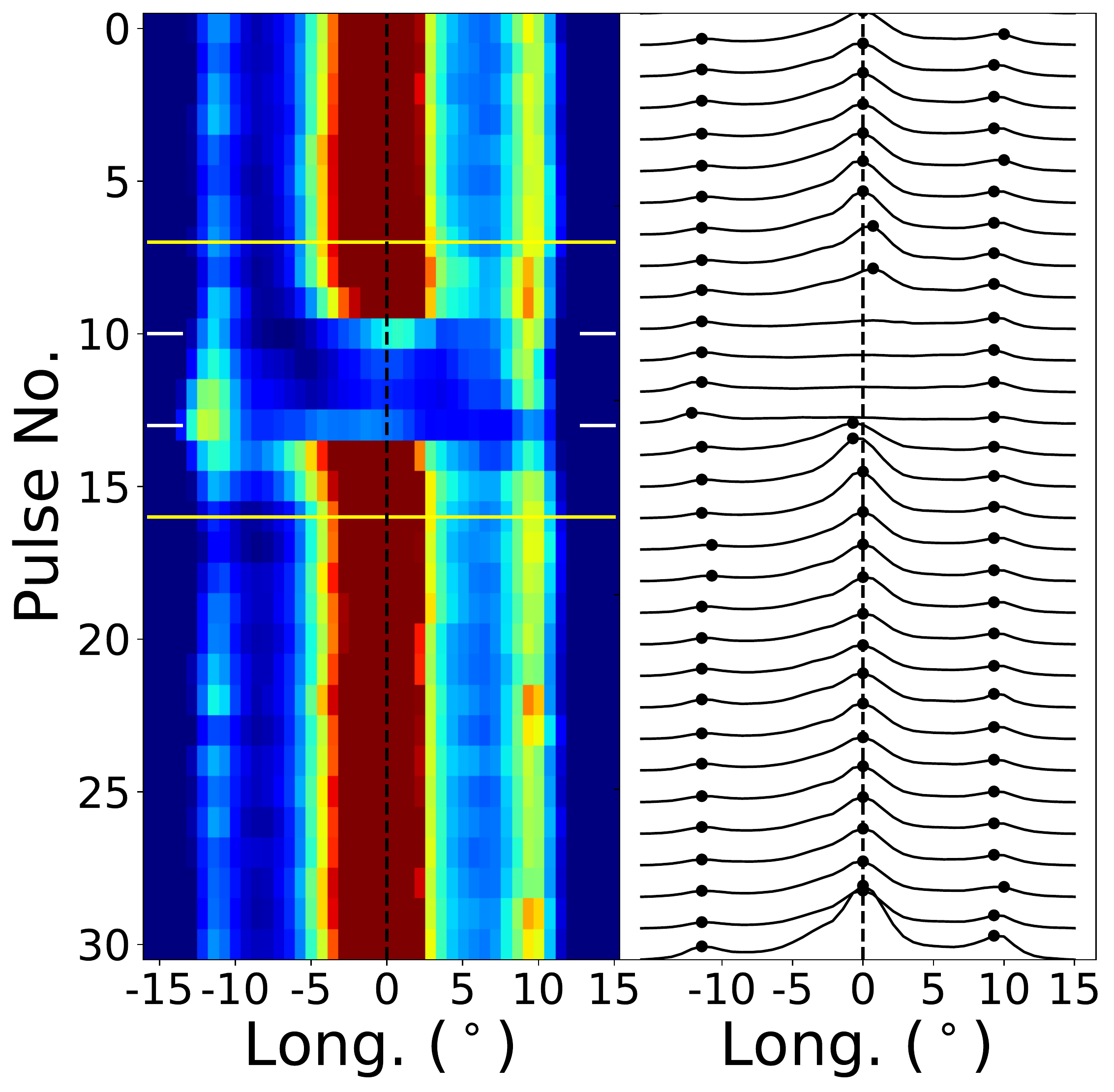}}
\subfigure[timescale=5]{
\label{fig2-c}
\includegraphics[width=0.32\linewidth]{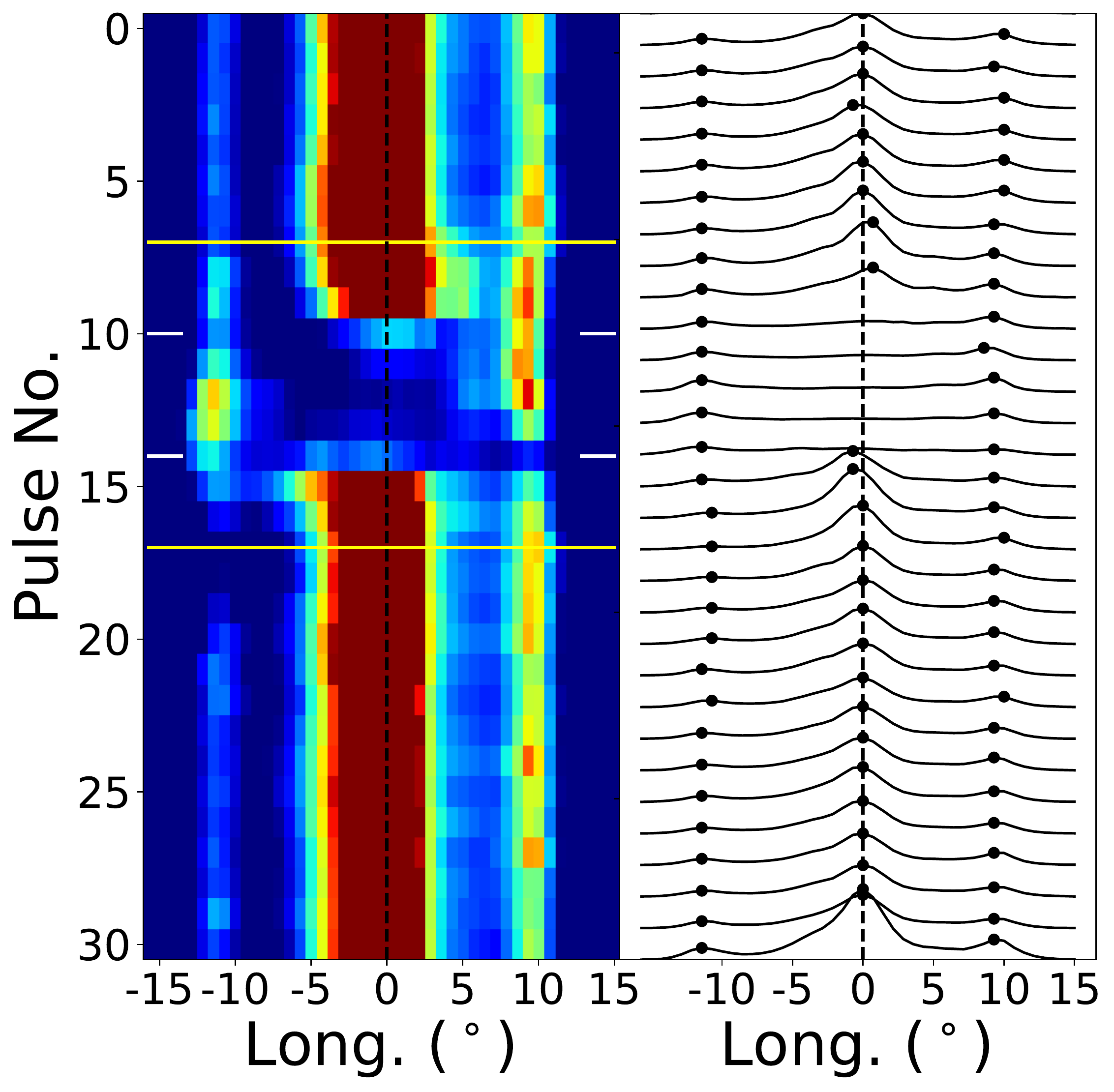}}
\subfigure[timescale=6]{
\label{fig2-d}
\includegraphics[width=0.32\linewidth]{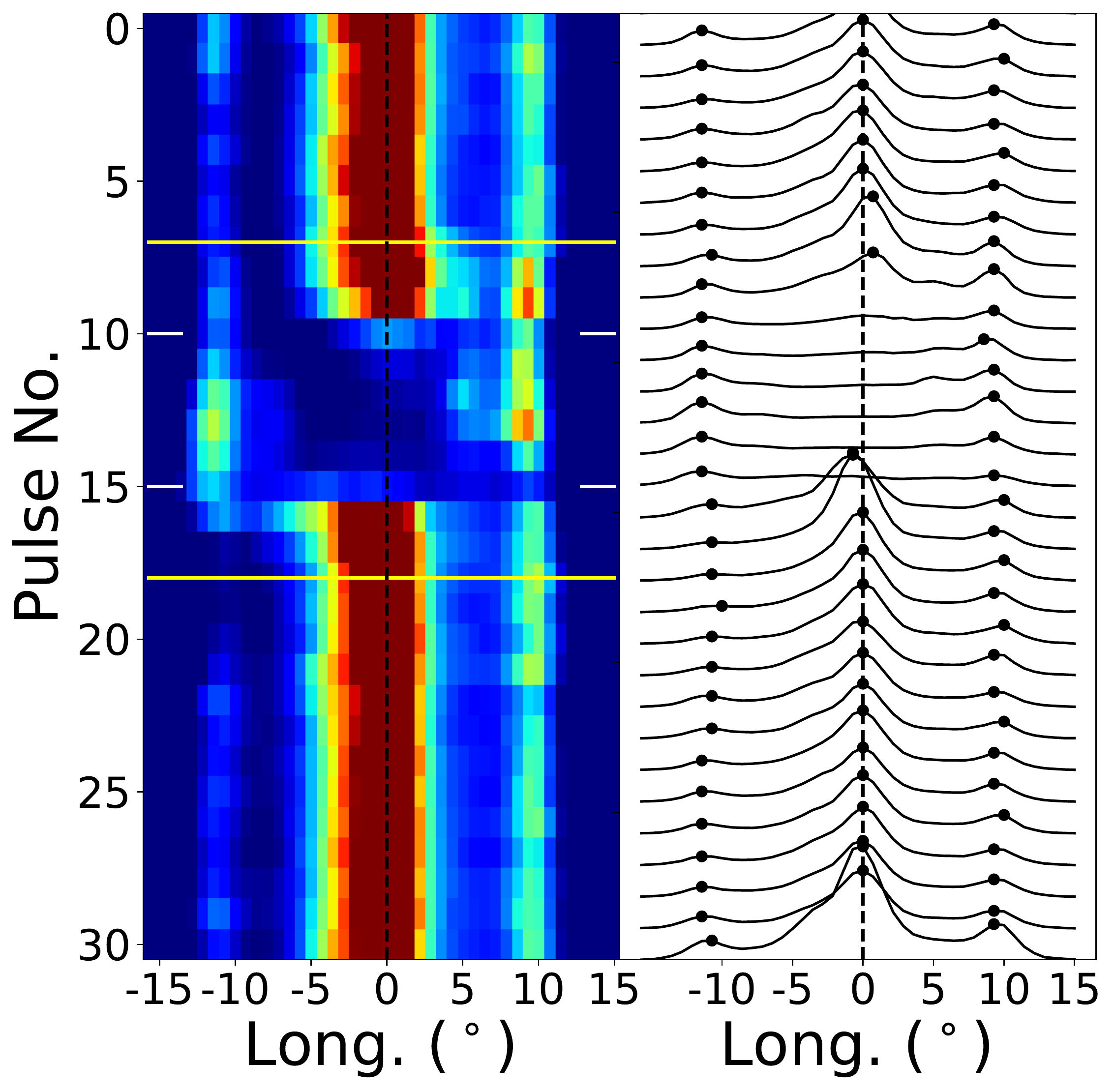}}
\subfigure[timescale=7]{
\label{fig2-e}
\includegraphics[width=0.32\linewidth]{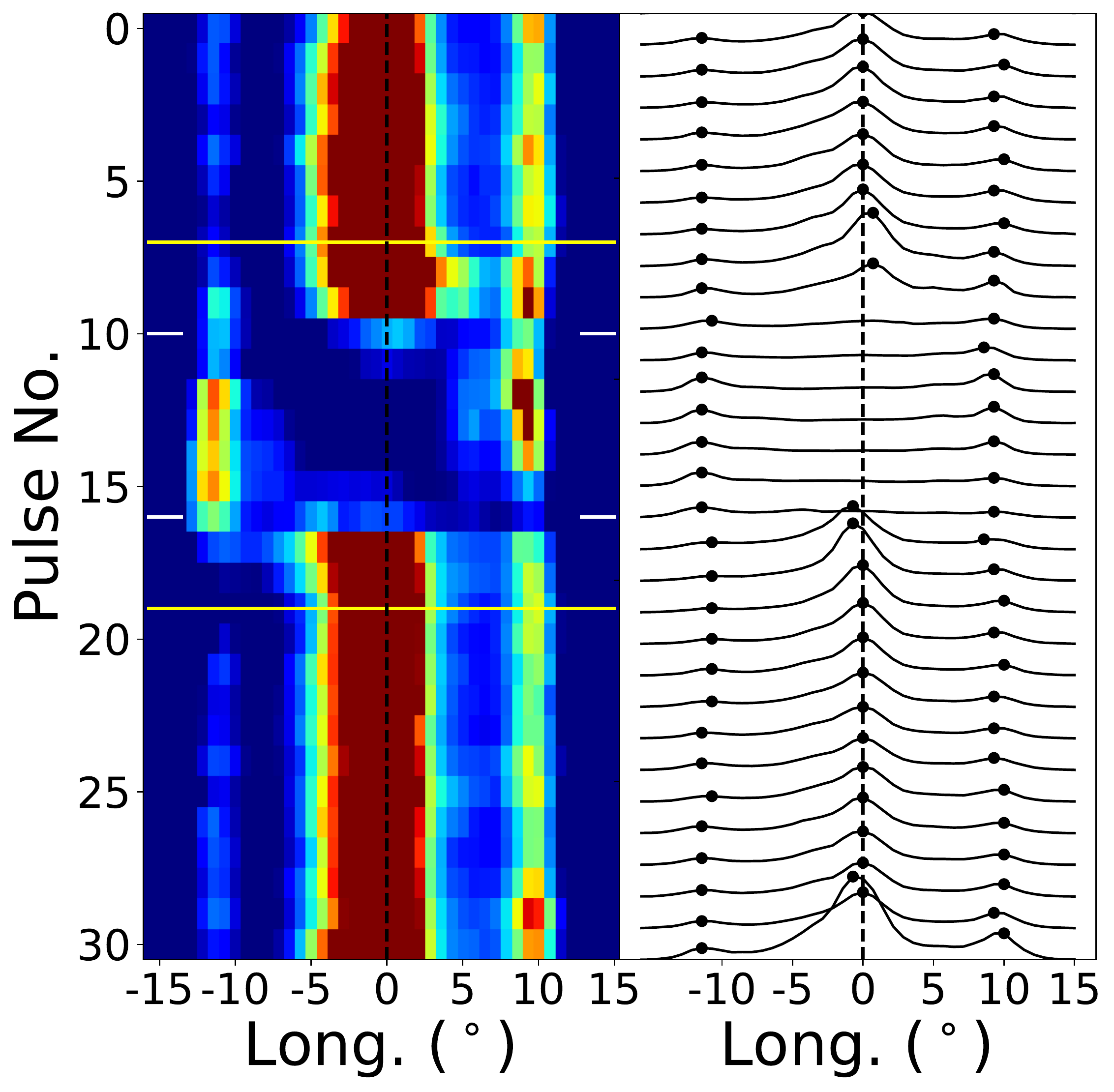}}
\subfigure[timescale=8]{
\label{fig2-f}
\includegraphics[width=0.32\linewidth]{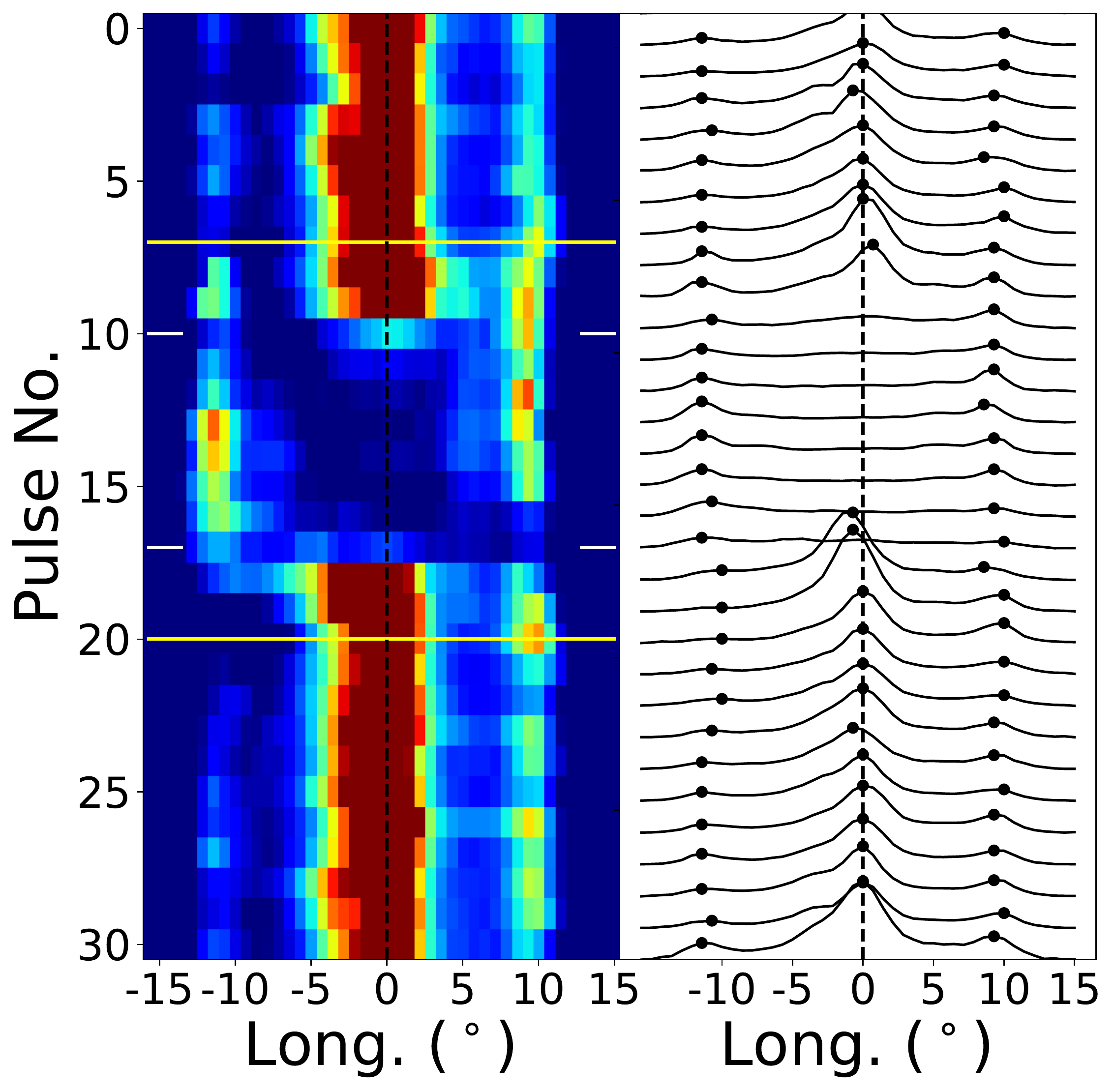}}
\subfigure[timescale=9]{
\label{fig2-g}
\includegraphics[width=0.32\linewidth]{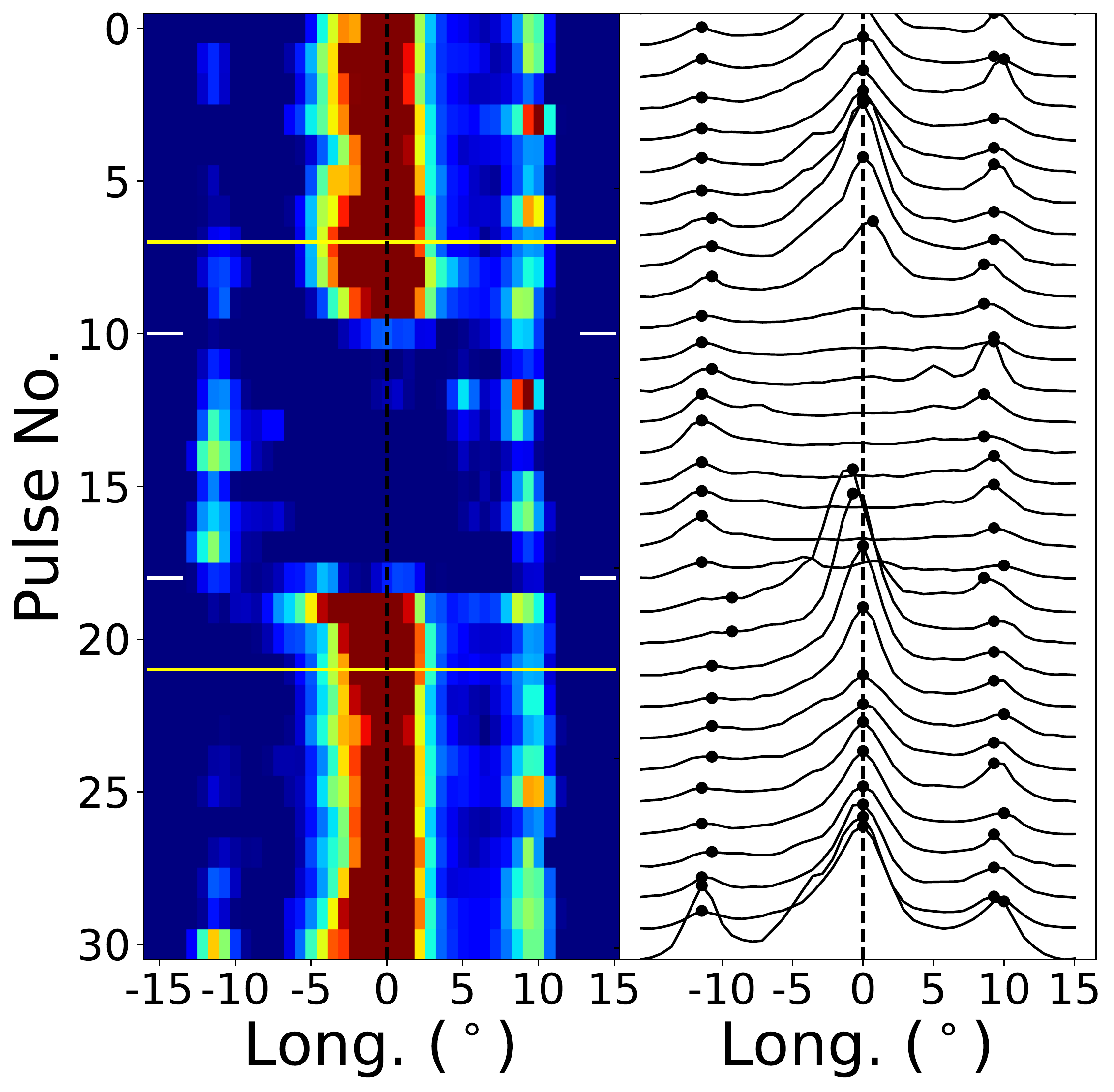}}
\subfigure[timescale=10]{
\label{fig2-h}
\includegraphics[width=0.32\linewidth]{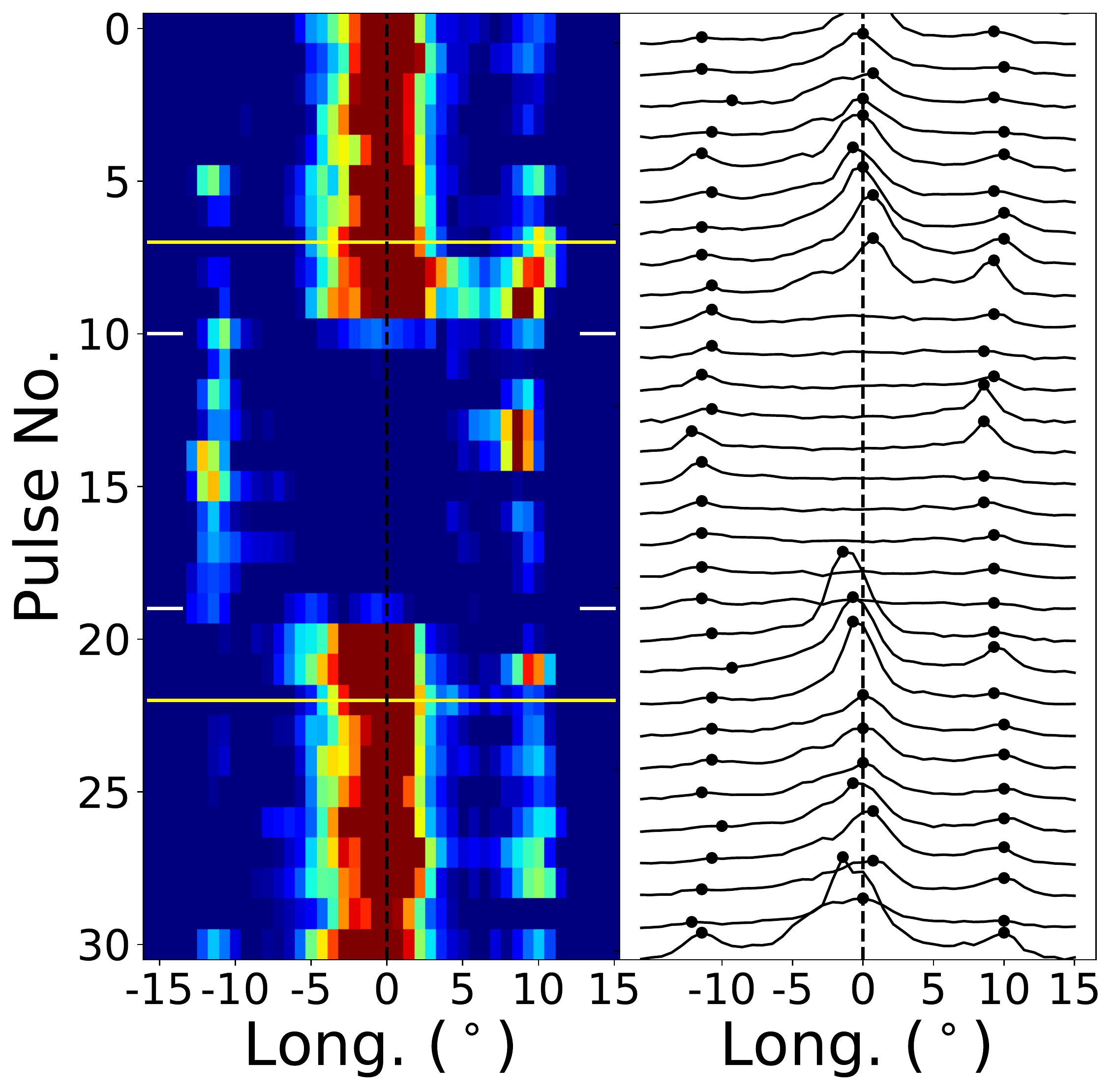}}
\caption{The distinct patterns are obtained by averaging many recognized emission patterns for a given time-scale. The left panel shows the distinct pattern indicated by the white lines. The recognized core-weak mode is in between the two short lines. The right panel shows the relevant single pulses which exhibit the variation of three main component peaks. The zero-longitude line is indicated by the black dashed line. Note that the peak value of core emission does not get down to the zero-baseline in all patterns (see  Figure~\ref{coreNullProfile}).}
\label{stablePatterns}
\end{figure*}

\begin{figure*}
\centering
\subfigure[the core component]{
\label{compareEVariance-b}
\includegraphics[width=0.45\linewidth]{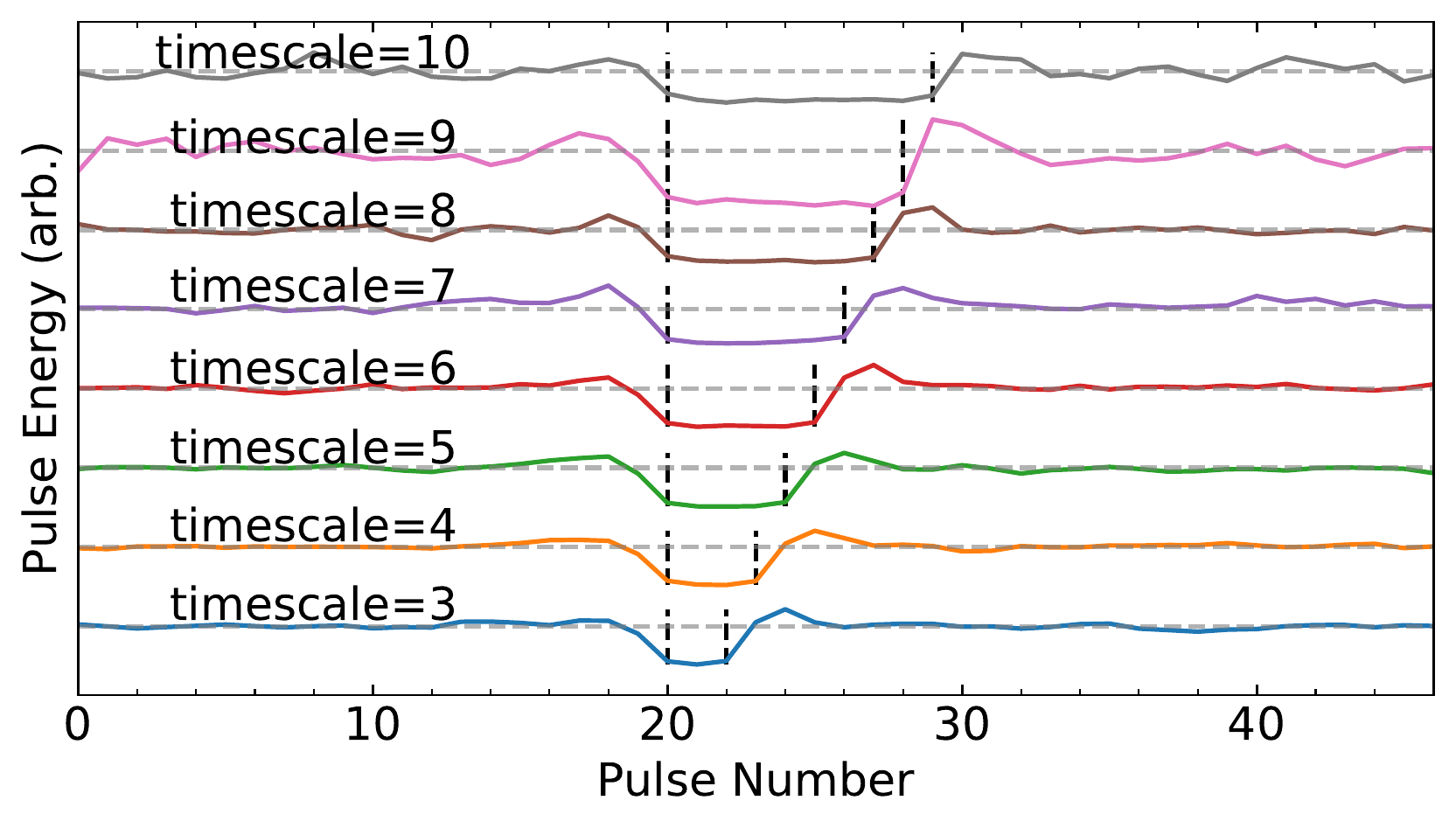}}
\subfigure[the leading component]{
\label{compareEVariance-a}
\includegraphics[width=0.45\linewidth]{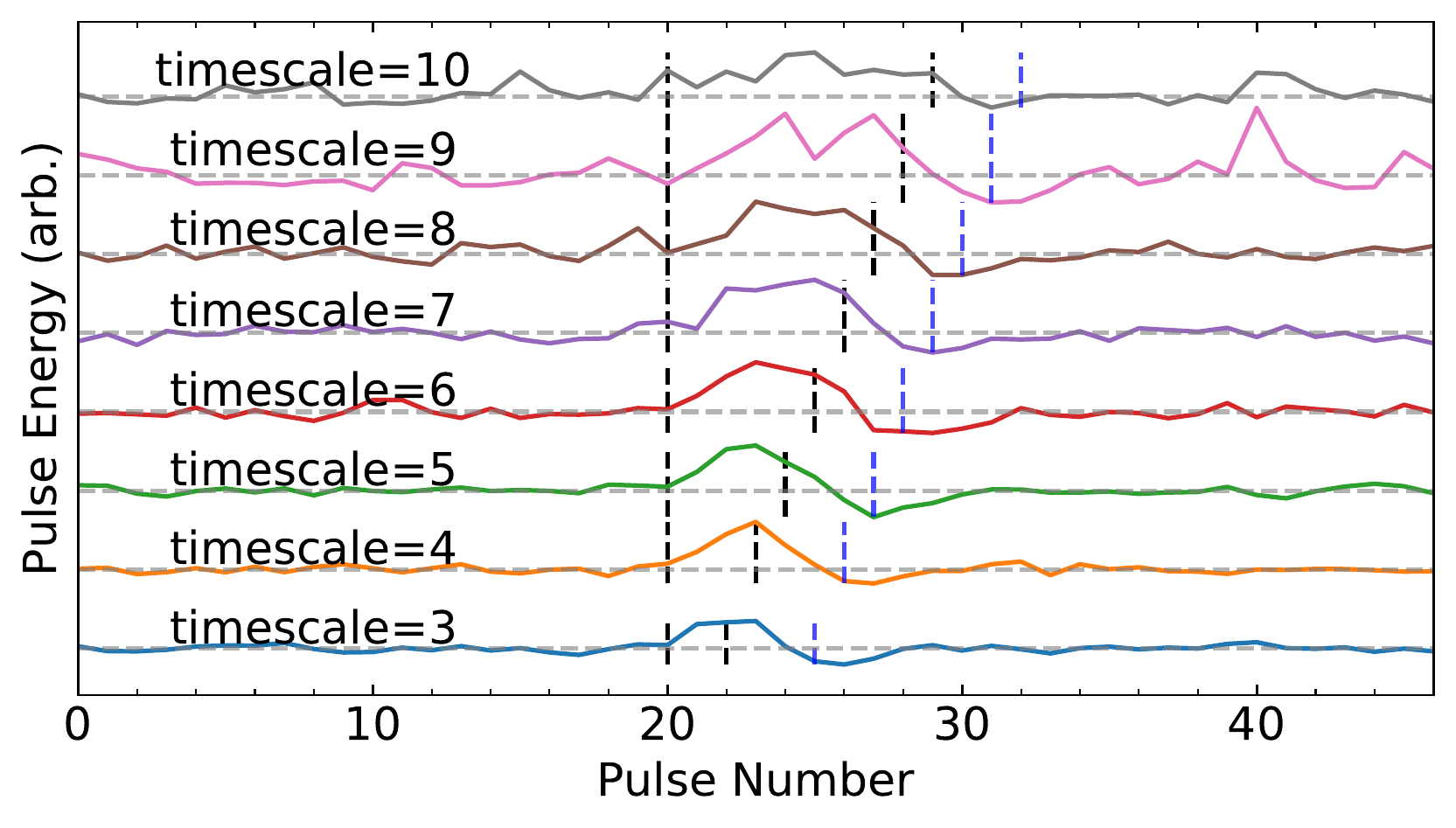}}
\subfigure[the trailing component]{
\label{compareEVariance-c}
\includegraphics[width=0.45\linewidth]{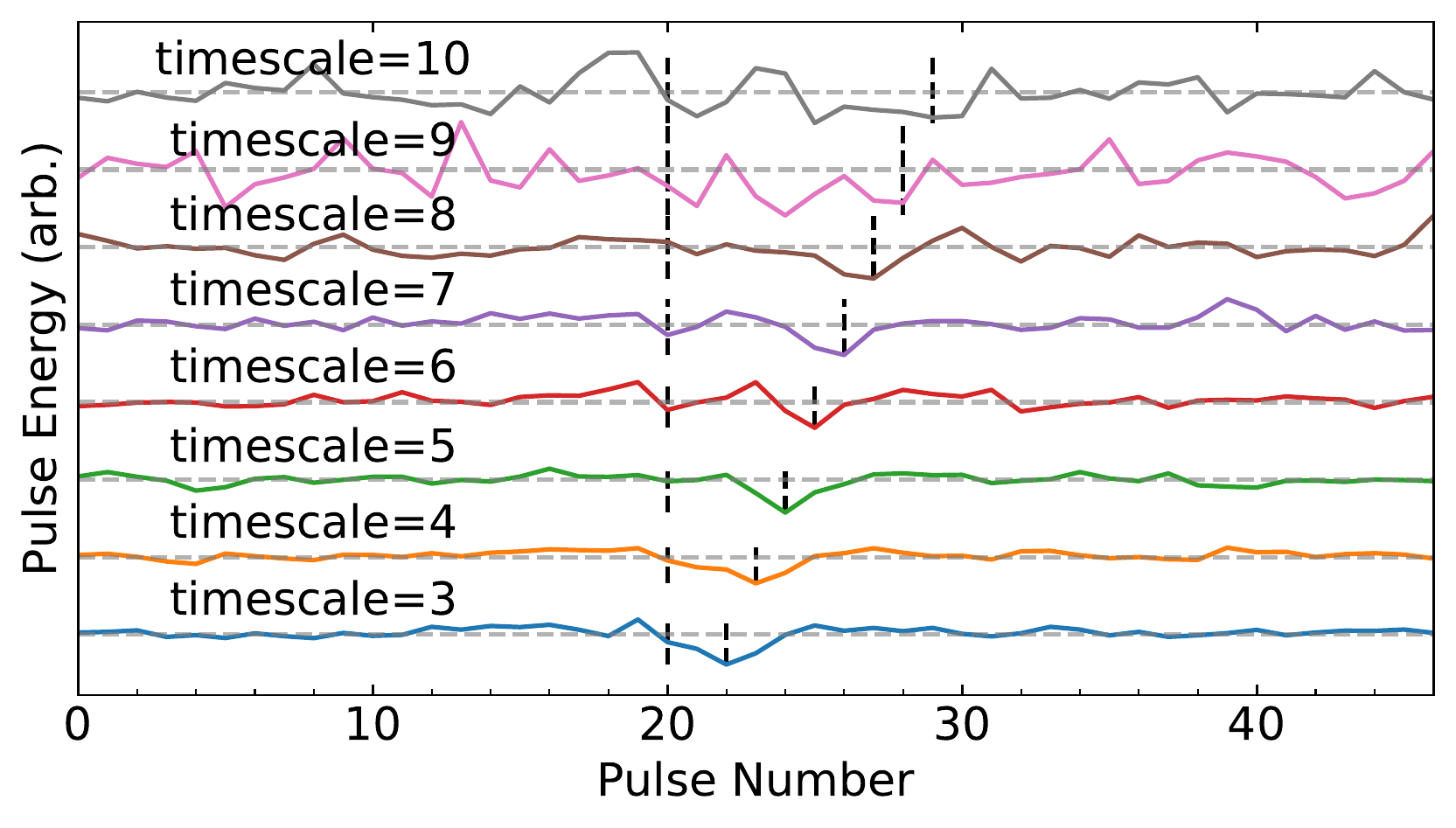}}
\subfigure[the whole single pulse]{
\label{compareEVariance-d}
\includegraphics[width=0.45\linewidth]{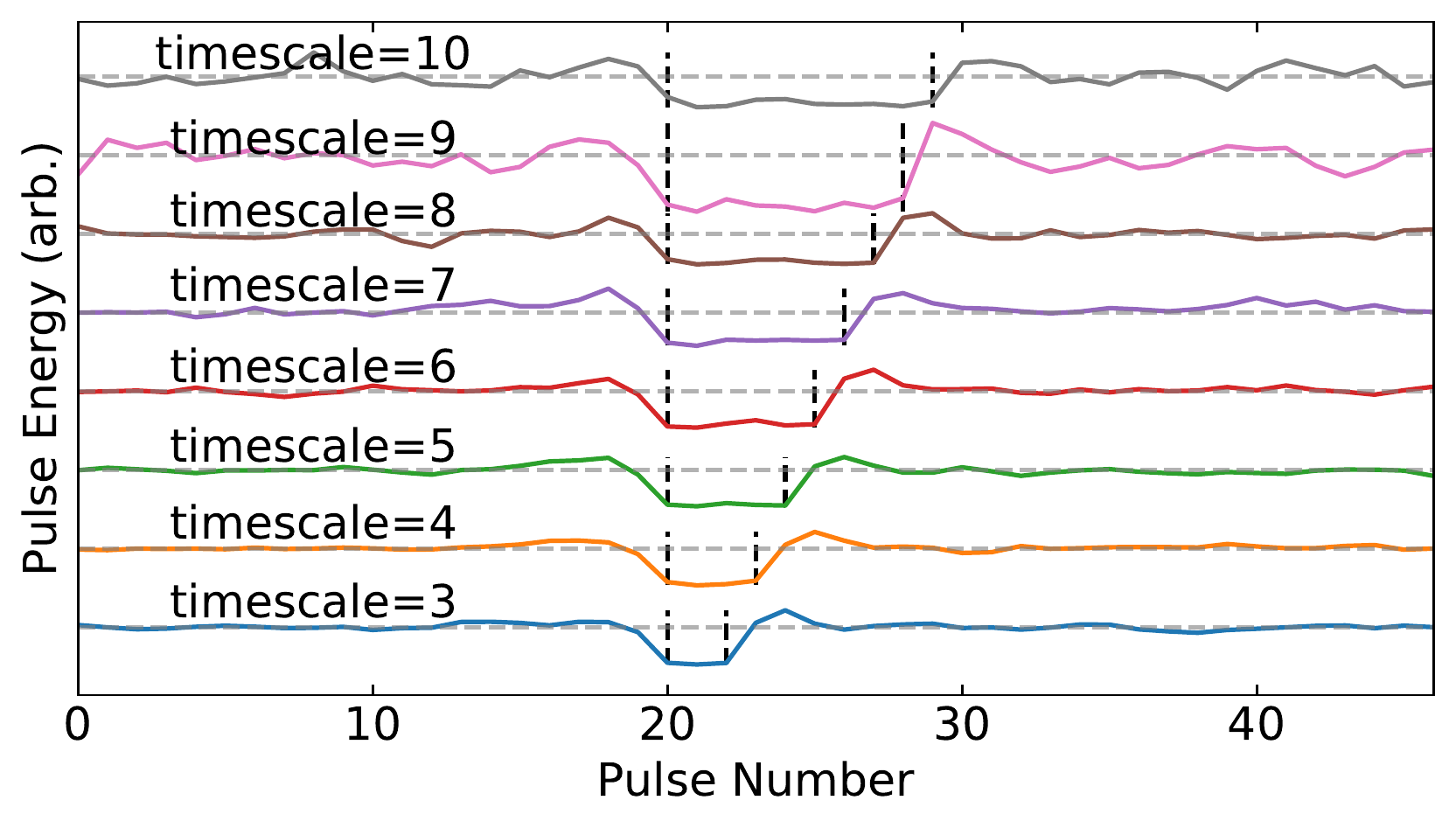}}
\caption{The pulse energy of the core, the leading and the trailing components as well as the whole individual pulse obtained from the averaged emission patterns in the phase ranges of $[-15^\circ, -8^\circ]$, $[-8^\circ, 7^\circ]$, $[7^\circ, 15^\circ]$ and $[-15^\circ, +15^\circ]$  in Figure~\ref{stablePatterns}, respectively. In addition to the periods for core-weak mode, the dips and bumps of energy variations for these components are intriguingly related and depend on timescale. The dashed lines represent the average amplitudes of normal sections in these patterns. The dotted vertical lines indicate the start and the end of core-weak mode. The third dashed line in panel (b) has a lag of 4 periods from the end of core-weak mode. 
}
\label{EnergyDis}
\end{figure*}

Through the statistics of core-weak patterns as the top panel of Figure~\ref{patternNumDis}, we find that the 5 observation sessions have similar distributions, which implies that there is no selection bias in the process for picking out patterns. Most of the patterns have a weak core lasting for 3 periods and the duration of longest core weak pattern can barely exceed 14 periods. The core-weak patterns occupy about 4\% of periods (see Figure~\ref{coreHist}), close to the result by \citet{tyw+2022} observed at the L-band. The distribution of total pattern numbers for all 5 observation sessions is shown in the bottom panel of Figure~\ref{patternNumDis}, and the positively skewed distribution can be fitted by a log-normal distribution function:
\begin{equation}
f(x) = \rm{A} \cdot \frac{1}{(2\pi)^{1/2}\sigma x} \cdot exp(\frac{(\rm{ln}(x)-\mu)^2}{-2\sigma^2}),
\end{equation}
here \rm{A} is the amplitude, $\mu$ is the mean value and $\sigma$ is the standard deviation. We get the parameters $\sigma=0.49\pm0.04$, $\mu=1.40\pm0.02$ and $\rm{A}=2662\pm324$. 

We check if the core-weak mode happens periodically. We set the middle single pulse with weak-core as "1", and all other single pulses as "0", and try the Fast Fourier Transform of the "1-0-1" sequences of all 4 observations sessions of PSR B0329+54 (except for the short observation at 2015). We get no evidences for the periodicity of weak core events. In other words, the core-weak pattern appears randomly.

For core-weak emission patterns with a given time scale, the patterns 
are added together, so that we get an averaged core-weak patterns, as shown in  Figure~\ref{stablePatterns} (a)-(h).

\subsection{The morphology of core-weak emission patterns}
\label{dynamicChange}

The core-weak emission patterns in Figure~\ref{stablePatterns} have 4 stages which can be depicted roughly as following.
 
(1) Stage No.1: before the beginning of the core-weak mode, the core emission slightly brightens (see Figure~\ref{EnergyDis} for pulse energy changes, i.e. integrated over the component), rather than directly decrease to a very low level. The peak first is shifted toward a later longitude phase, and the trailing component and the bridge between the core and trailing components become stronger than the normal situation.
 
\begin{figure*}
\centering
\subfigure[timescale=3]{
\label{fig3-a}
\includegraphics[width=0.24\linewidth]{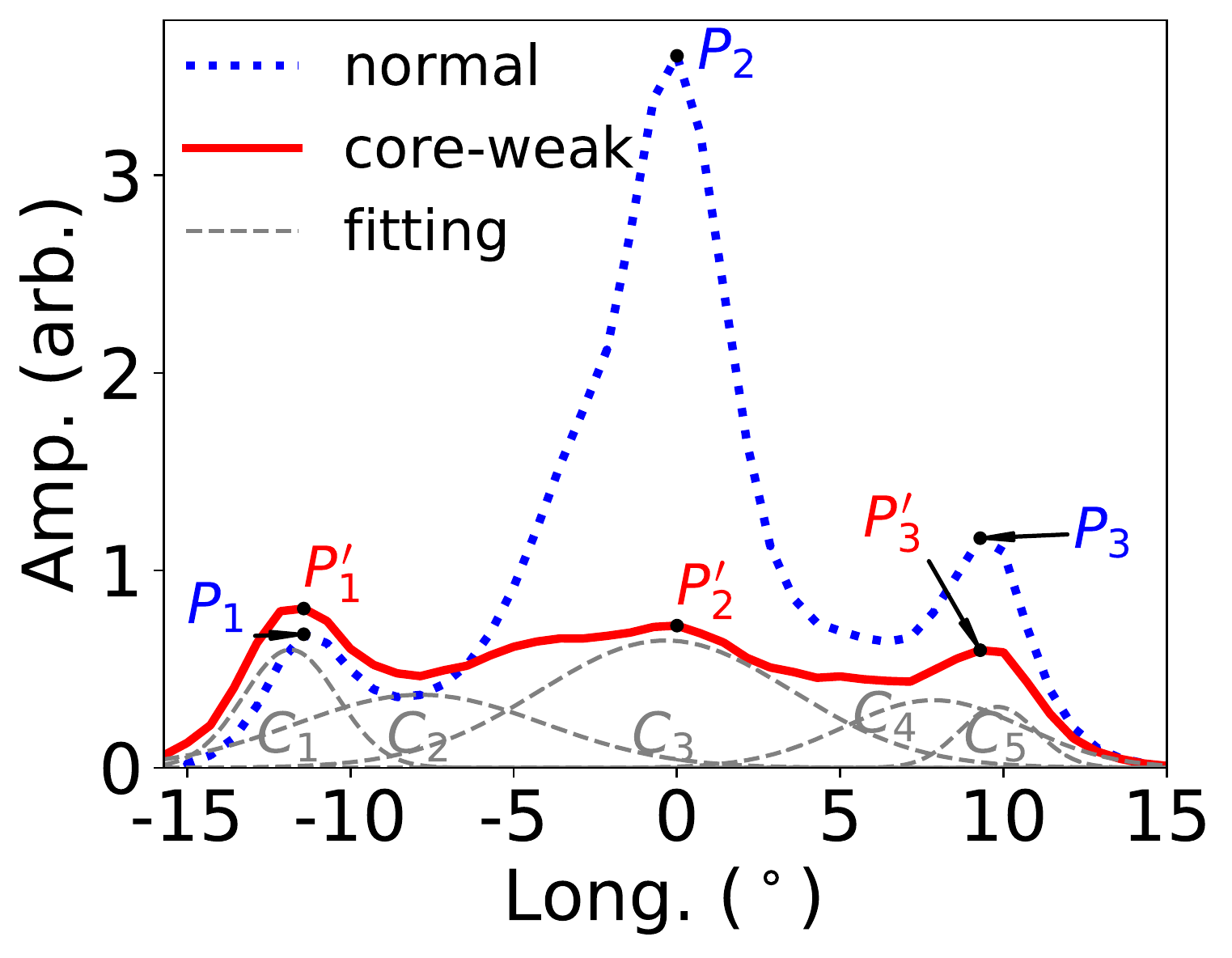}}
\subfigure[timescale=4]{
\label{fig3-b}
\includegraphics[width=0.24\linewidth]{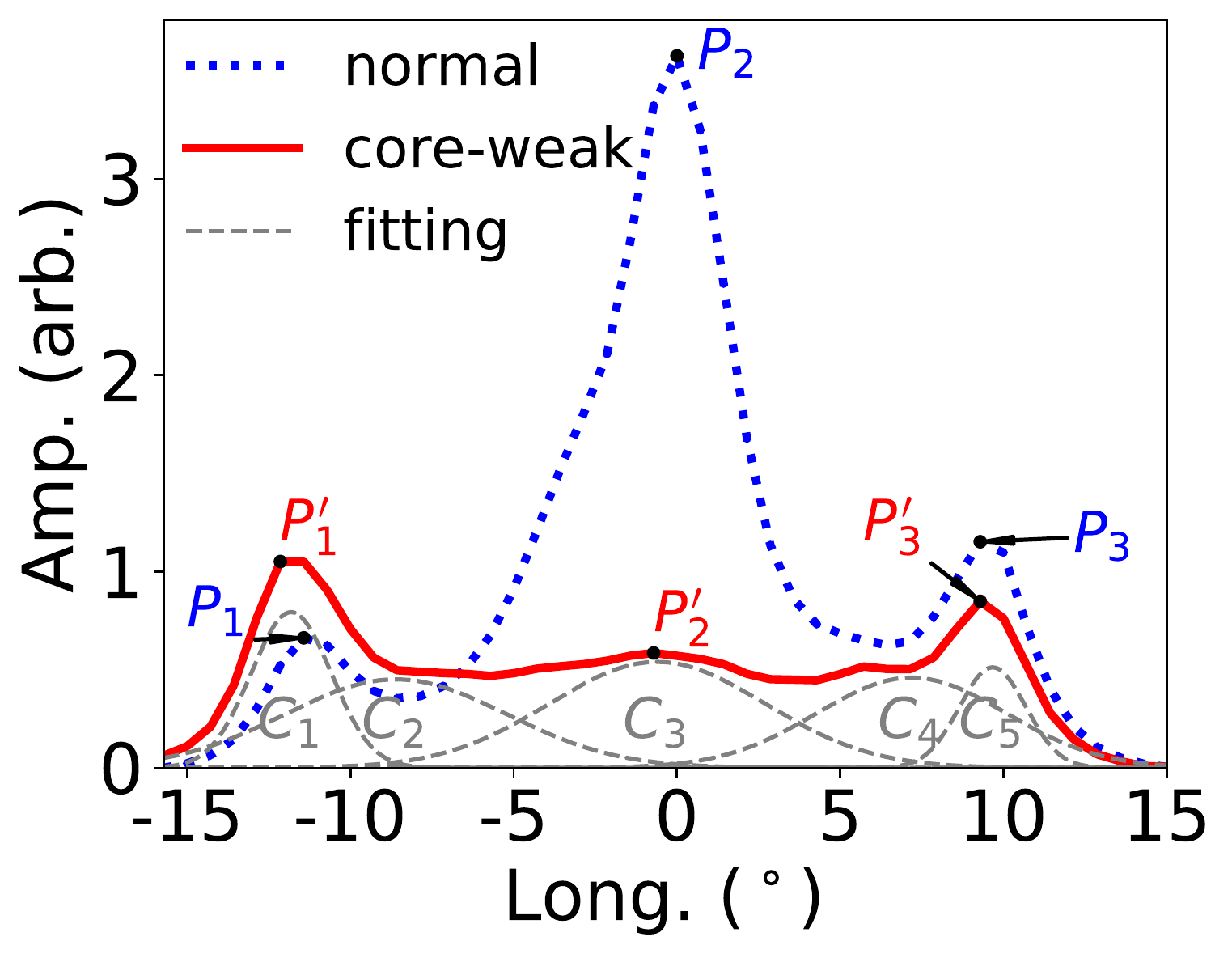}}
\subfigure[timescale=5]{
\label{fig3-c}
\includegraphics[width=0.24\linewidth]{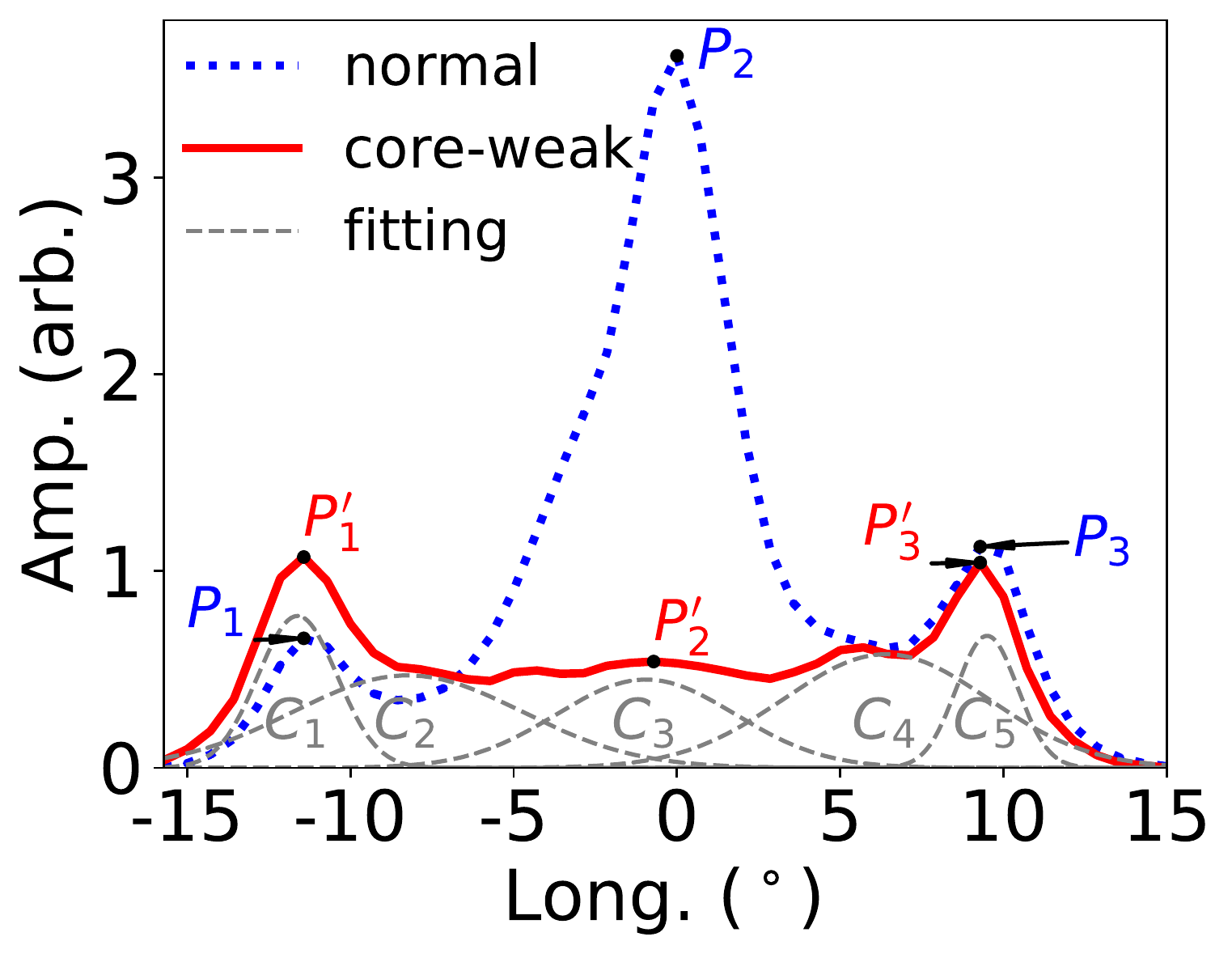}}
\subfigure[timescale=6]{
\label{fig3-d}
\includegraphics[width=0.24\linewidth]{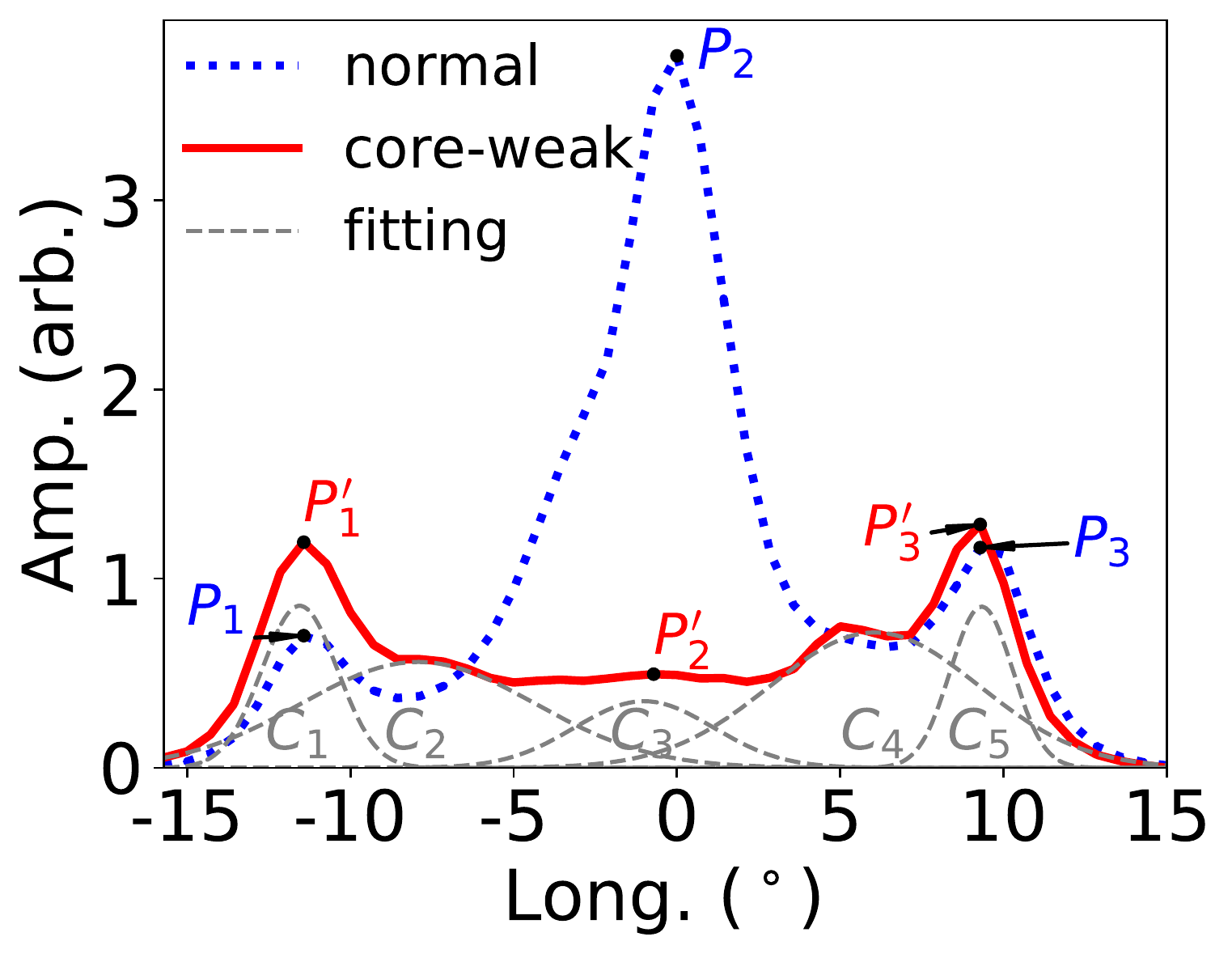}}
\subfigure[timescale=7]{
\label{fig3-e}
\includegraphics[width=0.24\linewidth]{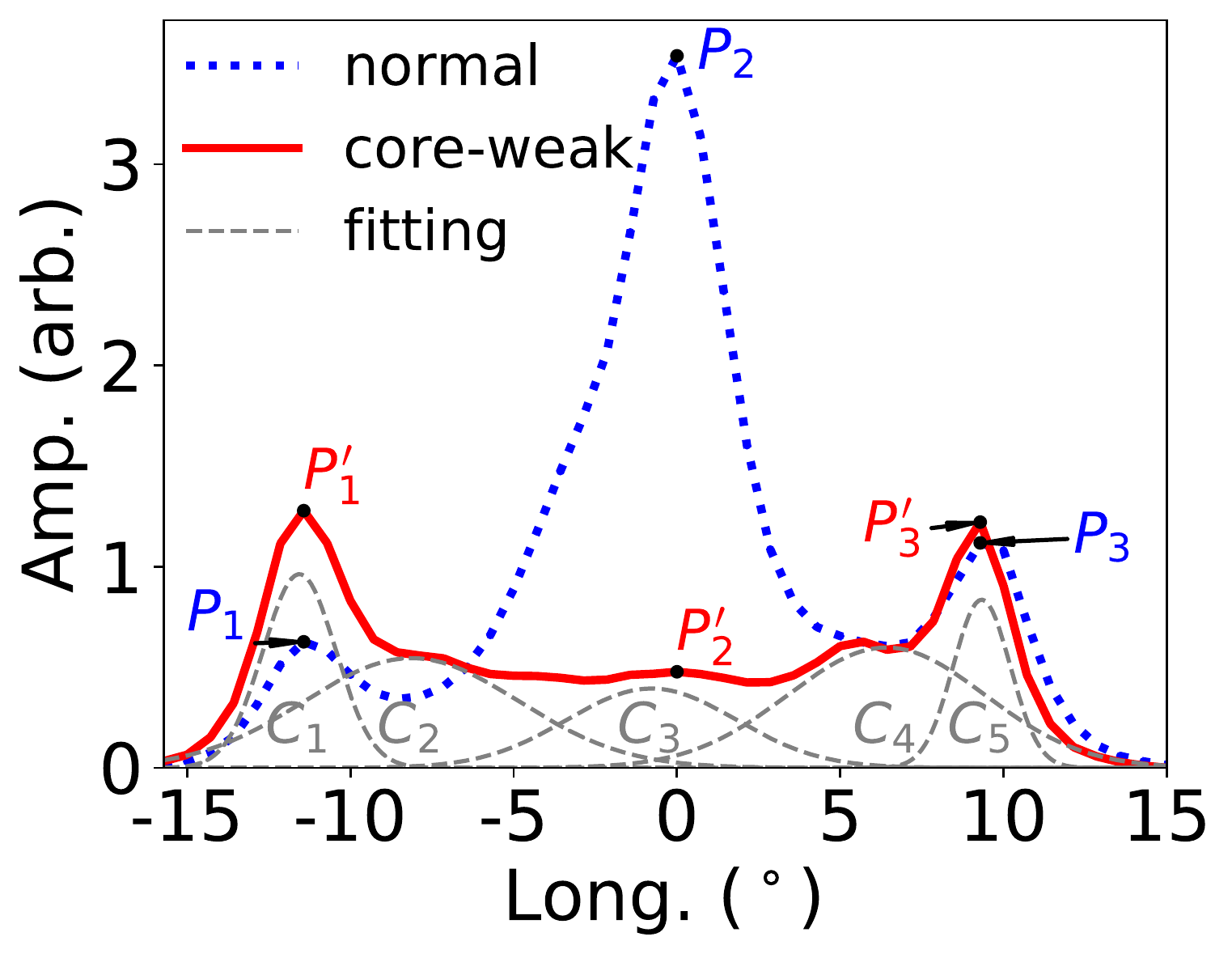}}
\subfigure[timescale=8]{
\label{fig3-f}
\includegraphics[width=0.24\linewidth]{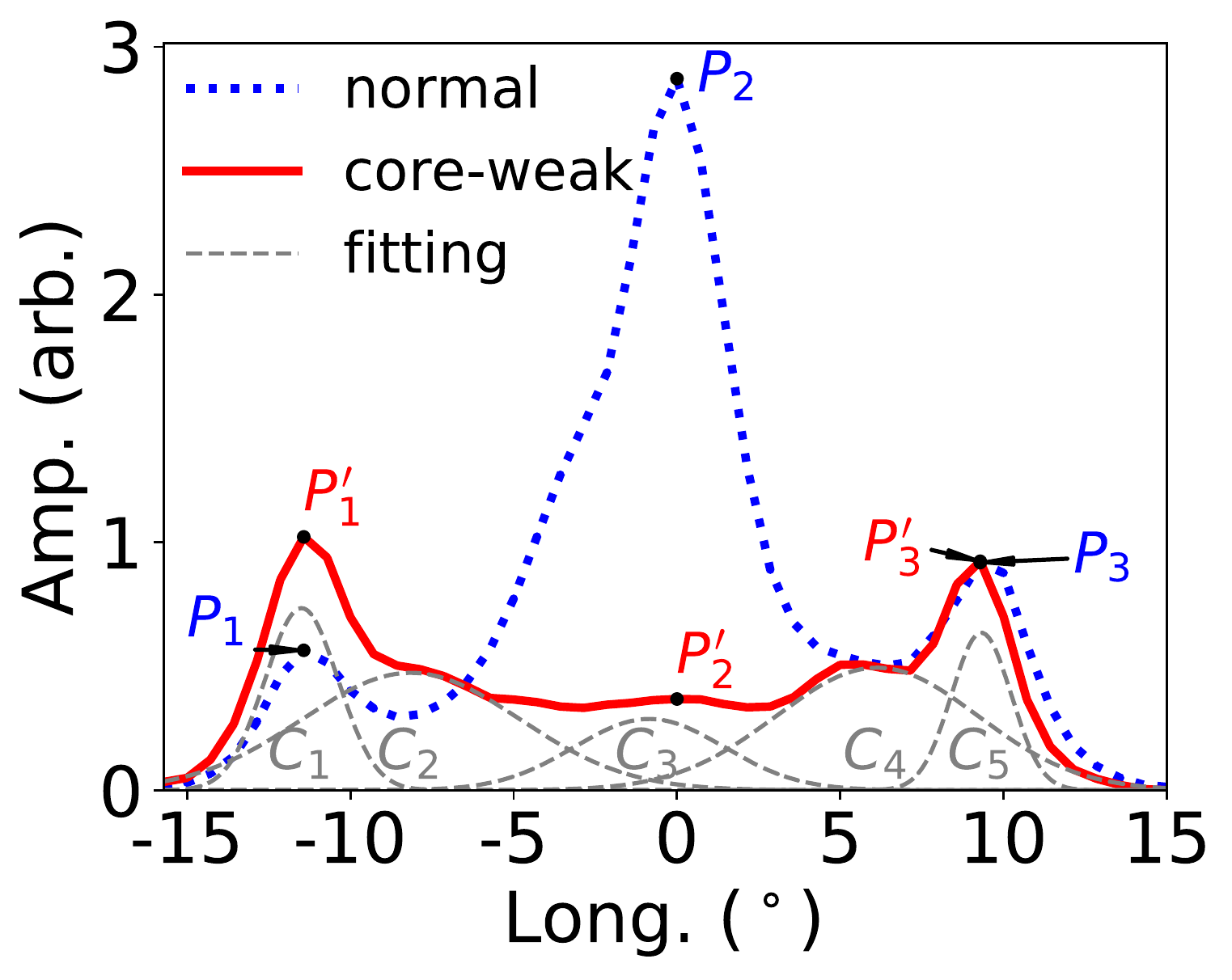}}
\subfigure[timescale=9]{
\label{fig3-g}
\includegraphics[width=0.24\linewidth]{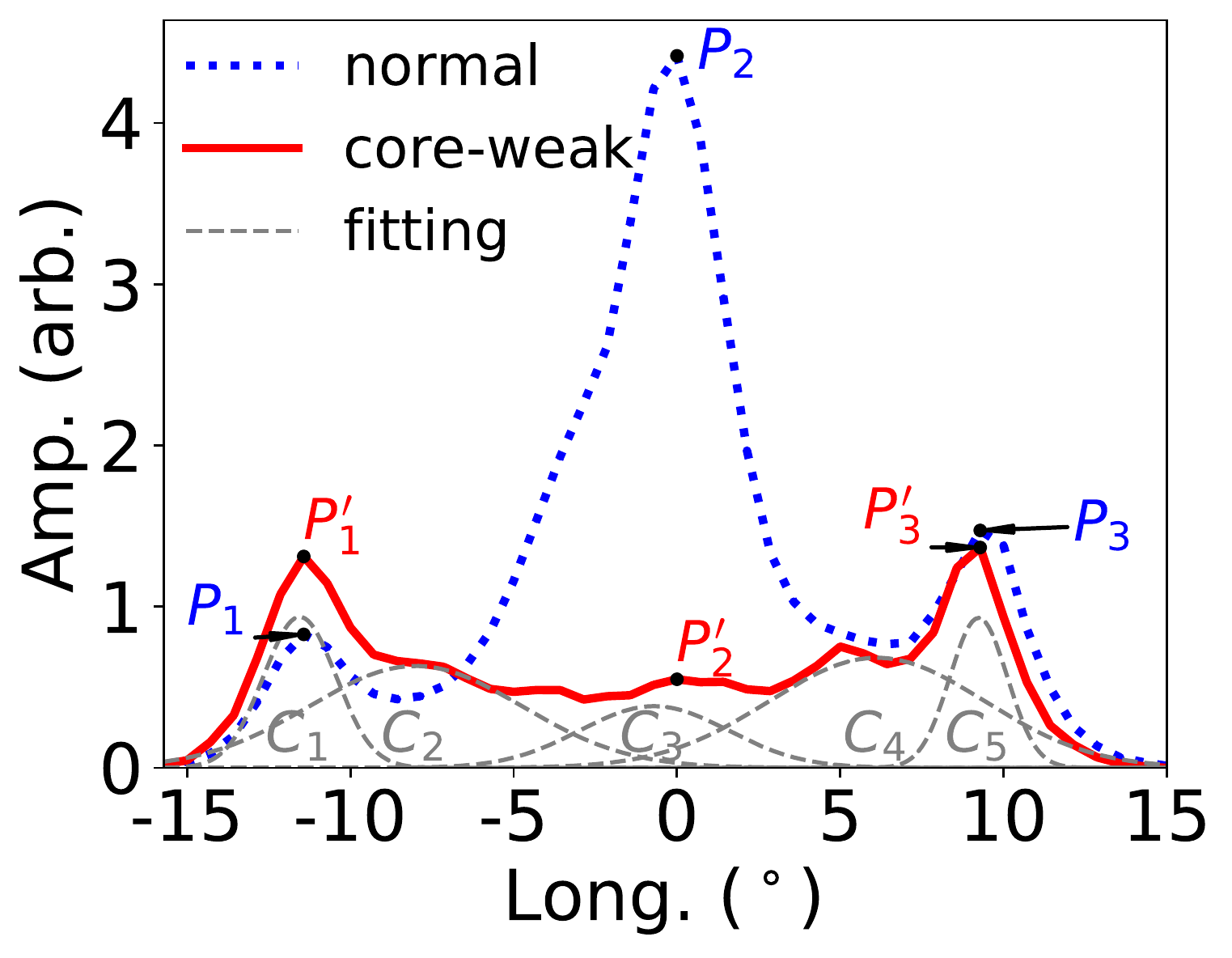}}
\subfigure[timescale=10]{
\label{fig3-h}
\includegraphics[width=0.24\linewidth]{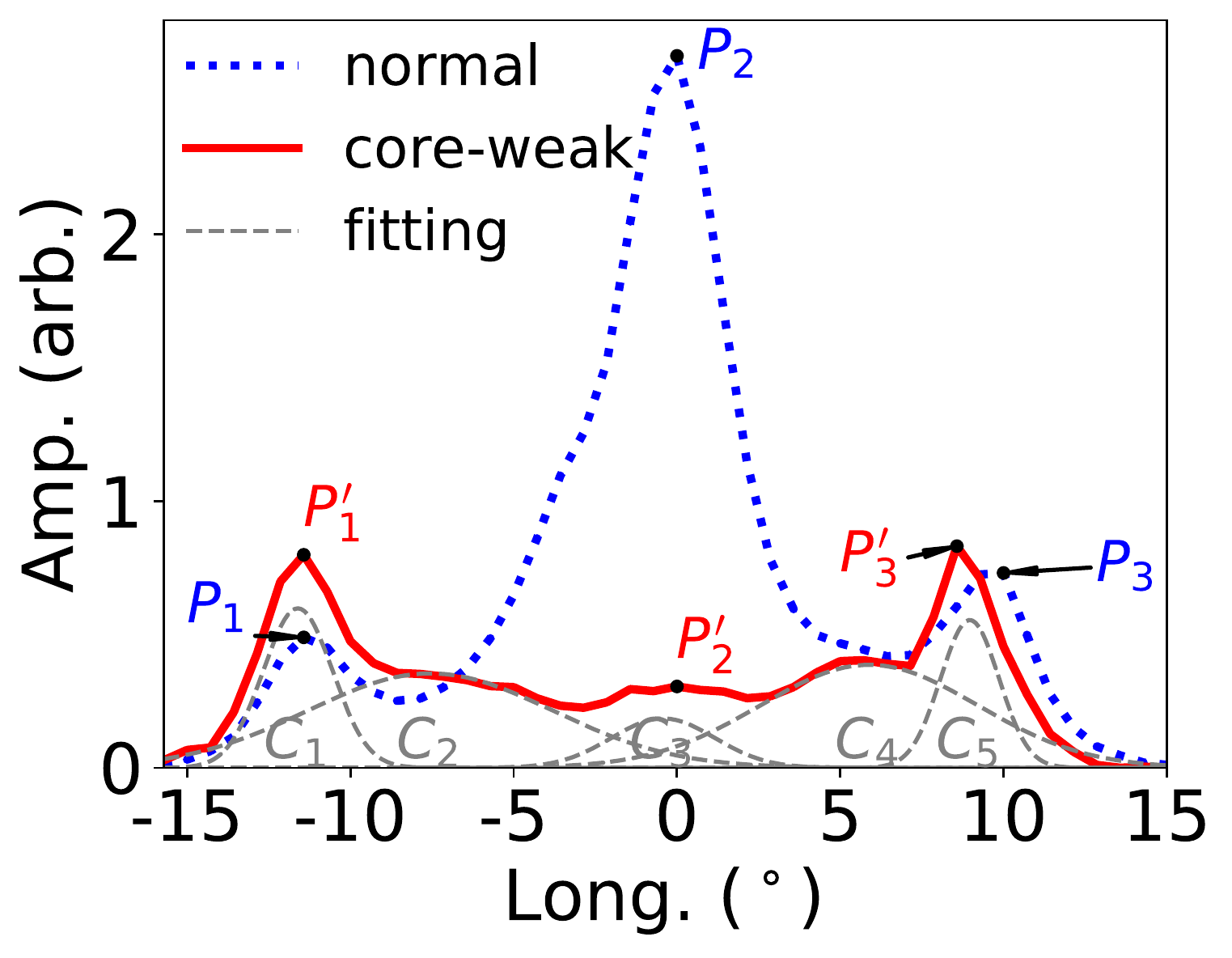}}
\caption{The averaged pulse profiles of core-weak mode for 8 different time-scales, compared with the profiles for the normal mode (dotted line). The core-weak profiles can be fitted with 5 Gaussian components as shown by dashed lines. The peaks of three prominent components in the normal mean pulse are labeled as $P_1$, $P_2$ and $P_3$, and peaks for core-weak modes are labeled as $P_1'$, $P_2'$ and $P_3'$. The 5 fitted Gaussian components are labeled with $C_1$, $C_2$, $C_3$, $C_4$ and $C_5$. }
\label{coreNullProfile}
\end{figure*}

(2) Stage No.2: the core becomes really weak, probably appearing as ``null'' in a low sensitivity observations \citep{MitraD2007_MNRAS}. The leading component becomes gradually stronger. In addition, two weak components become enhanced in the two bridge phases (see Figure~\ref{coreNullProfile} for components). Conversely, the trailing component and the bridge between the core and trailing components get weak gradually but not to the zero level at the S-band we observed. The trailing component reaches its intensity minimum at the end of core-weak mode. 

(3) Stage No.3: the core emission starts to recover at an earlier phase than the normal core peak (see Figure~\ref{EnergyDis} for details). Subsequently, the core component gets the intensity enhanced and then goes back to the normal state. Similarly, the trailing component goes back to the normal state in a few periods and the leading component continuously get weak for some periods.
 
(4) Stage No.4: after both the core and trailing component return to the normal state, the leading component tends to get weaker and then slowly recovers to its normal amplitude in 3 or 4 periods, accompanied with a slight phase shift toward a larger longitude phase.
 
Such relatedly variations of subpulse components occur in all patterns with 8 distinct time-scales.

\subsection{Variation of pulse components in the core-weak patterns}
\label{FluxVariation}

As discussed above, the core-weak emission patterns show relatedly variations of profile components in the phase-time plot. The pulse energy variation of the leading, the core and the trailing components as well as the whole pulse are plotted along time (in period numbers) in Figure~\ref{EnergyDis}.

For the core component, the pulse energy declines rapidly to a low level after an enhancement, and comes back to a normal state with another enhancement, see  Figure~\ref{EnergyDis}. The duration of enhancements depends on the time scale of core-weak mode. 

For the leading component, the pulse energy starts to increase first and does not return to normal even after the core-weak mode. It starts to decreases after the core-weak mode, and get a dip after generally 4 periods, as showed in Figure~\ref{EnergyDis}b. In other words, the energy of the leading components are enhanced during the core-weak pattern and has a dip with a 4 period lag afterwards.  

The phase range of the trailing component we set covers the bridge between the central core and the dominant trailing peak. For the trailing component, the pulse energy starts to decrease when a core-weak pattern starts, mainly due to the weakened bridge. It continuously decreases to the minimum at the end of core-weak mode. The trailing peak itself declines to the minimum in a short time. 

The total energy of individual pulses for all the core-weak patterns show a slight enhancements both prior and post to the core-weak mode, and it stays at a constantly lower level during the core-weak mode.

\begin{table}
    \caption[]{The relative peak amplitude changes of the 3 prominent component marked in Figure~\ref{coreNullProfile} for the core-weak patterns of different timescales}
    \label{phaseChange}
    \centering
    \begin{tabular}{cccc}
    \hline
	Timescales~& $P_{1}'/P_{1}$ & $P_{2}'/P_{2}$ & $P_{3}'/P_{3}$ \\ 
	(in~period) &  &  &  \\
    \hline
	3&1.19 &0.20 &0.51\\
	4&1.59 &0.16 &0.74\\
	5&1.63 &0.15 &0.93\\
	6&1.71 &0.13 &1.10\\
	7&2.04 &0.14 &1.09\\
	8&1.81 &0.13 &1.01\\
	9&1.58 &0.12 &0.93\\
	10&1.63 &0.11 &1.14\\ \hline 
	\rm{Mean} & 1.65 & 0.14 & 0.93\\
    \hline
    \end{tabular}
\end{table}

\begin{table*}
    \caption[]{The longitude locations and the half-width of 5 fitted Gaussian components to the averaged profiles for core-weak mode with 8 timescales. }
    \label{5GaussianFitting}
    \centering
    \begin{tabular}{ccccccccccccc}
    \hline
    \rm{Timescales} & Longitude$_1$ & Longitude$_2$ & Longitude$_3$ & Longitude$_4$ & Longitude$_5$ \\ 
    \rm{(in~period)}  & $(^\circ)$ & $(^\circ)$ & $(^\circ)$ & $(^\circ)$ & $(^\circ)$\\ 
    \hline
	3 & -11.65$\pm$0.05 & -7.73$\pm$2.97 & -0.34$\pm$1.25 & 7.73$\pm$0.88 & 9.67$\pm$0.09 \\
	4 & -11.63$\pm$0.03 & -8.47$\pm$0.59 & -0.60$\pm$0.22  & 7.07$\pm$0.39 & 9.51$\pm$0.03 \\
	5 & -11.49$\pm$0.06 & -9.14$\pm$1.59 & -1.02$\pm$0.34 & 5.81$\pm$0.11 & 9.23$\pm$0.04 \\
	6 & -11.41$\pm$0.03 & -9.14$\pm$0.20 &-0.13$\pm$0.43  & 5.49$\pm$0.07 & 9.04$\pm$0.02 \\
	7 & -11.38$\pm$0.02 & -8.00$\pm$0.31  &-0.77$\pm$0.21  & 6.31$\pm$0.23 & 9.19$\pm$0.02 \\
	8 & -11.34$\pm$0.02 & -9.08$\pm$0.15 & 0.12$\pm$0.36  & 5.65$\pm$0.08 & 9.07$\pm$0.02 \\
	9 & -11.41$\pm$0.04 & -8.86$\pm$0.28 & 1.04$\pm$0.49  & 5.44$\pm$0.10  & 8.98$\pm$0.03 \\
	10 & -11.45$\pm$0.05 & -7.83$\pm$0.45 &0.78$\pm$0.35  & 5.22$\pm$0.14 & 8.72$\pm$0.06 \\
	
	\hline
	\rm{Mean}  & -11.47 & -8.53 & -0.11 & 6.09 & 9.18 \\
    \hline
    
	\hline
    \rm{Timescales} & $w_1$ & $w_2$ & $w_3$ & $w_4$ & $w_5$ & \\ 
    \rm{(in ~period)}  & $(^\circ)$ & $(^\circ)$ & $(^\circ)$ & $(^\circ)$ & $(^\circ)$\\ 

    \hline
    3 & 1.67$\pm$0.07 & 4.39$\pm$0.97 & 4.47$\pm$1.35 & 4.17$\pm$0.37 & 1.36$\pm$0.12 \\
    4 & 1.42$\pm$0.04 & 3.93$\pm$0.27 & 3.82$\pm$0.47 & 3.35$\pm$0.17 & 1.18$\pm$0.04 \\
    5 & 1.41$\pm$0.05 & 4.07$\pm$0.32 & 3.17$\pm$0.41 & 3.63$\pm$0.17 & 1.08$\pm$0.04 \\
    6 & 1.34$\pm$0.04 & 4.15$\pm$0.21 & 2.46$\pm$0.25 & 3.70$\pm$0.12 & 1.10$\pm$0.03 \\
    7 & 1.29$\pm$0.03 & 3.81$\pm$0.17 & 3.00$\pm$0.31 & 3.50$\pm$0.13 & 1.01$\pm$0.03 \\
    8 & 1.30$\pm$0.03 & 3.85$\pm$0.15 & 2.65$\pm$0.24 & 3.58$\pm$0.13 & 1.06$\pm$0.03 \\
    9 & 1.29$\pm$0.07 & 3.77$\pm$0.28 & 2.63$\pm$0.45 & 3.82$\pm$0.26 & 1.02$\pm$0.05 \\
    10 & 1.23$\pm$0.06 & 4.34$\pm$0.27 & 1.83$\pm$0.27 & 3.86$\pm$0.21 & 1.04$\pm$0.05 \\
    \hline
    \rm{Mean} & 1.37 & 4.04 & 3.00 & 3.70 & 1.11 \\
    \hline
    \end{tabular}
\end{table*}

\subsection{Averaged profiles for core-weak modes}
\label{profile}

By integrating individual pulses in the core-weak mode with different timescales, we get 8 abnormal average profiles as shown in Figure~\ref{coreNullProfile}. The mean profiles of normal pulses are plotted for comparison. The peaks of three prominent components of the normal profile are labeled as $P_1$, $P_2$ and $P_3$, and those of core-weak profiles are labeled as $P_1'$, $P_2'$ and $P_3'$. The ratios of $P_1'/P_1$, $P_2'/P_2$ and $P_3'/P_3$ for core-weak modes with 8 timescales are listed in Table~\ref{phaseChange}. The core component during the core-weak patterns decrease to about $14\%$ of the normal state, while the leading component is enhanced to about $165\%$ of the normal state, and the trailing component does not change much for core-weak modes with long time-scales but does decrease to $50\%$ for the modes with a short time-scale. 

When the core component gets weak, two bridge components in the phase ranges of $[-9^\circ,-6^\circ]$ and $[4^\circ,7^\circ]$ become relatively stronger, and their peaks can be marginally distinguishable in Figure~\ref{coreNullProfile}. We can fit the abnormal core-weak profiles with 5 Gaussian components \citep[see e.g.][]{ KramerM1994_AA}, $C_1$, $C_2$, $C_3$, $C_4$ and $C_5$. The phase-longitude and width are listed in Table~\ref{5GaussianFitting}. Obviously, the rear Gaussian components are closer than the precedent Gaussian components, which can be explained by the aberration and retardation effect \citep{Gangadhara2001}.

The so-called `pedestal' emission component \citep{MitraD2007_MNRAS} has been detected precede to the core at around $-3^\circ$, though it is more obvious at previously low frequency observations and seems to be independent with the core component \citep{MitraD2007_MNRAS}. When the core component gets weak, the  `pedestal' emission component decreases to disappearance in Figure~\ref{stablePatterns}, and the whole process takes about 3 periods. Then, the  `pedestal' emission component appears again about 3 periods after the core component starts a recovery.

\section{Summary and Discussion}
\label{DiscussionAndConclusion}

We carried out 5 long observations of PSR B0329+54 by using Jiamusi 66-m telescope at 2.25 GHz from 2015 to 2017. From the series of individual pulses, we recognized 8 mode-changing events. We identified the core-weak mode with different time-scales, which was first noticed as `core-null' events by \citet{MitraD2007_MNRAS} and recently studied by \citet{tyw+2022}. The total period number for the core-weak mode occur about 4\% of periods at the S-band we observed, close to 3.6\% obtained by  \citet{tyw+2022} for 1.5~GHz data. We find no periodicity for the core-weak mode occurring, but the event-rates follow a log-normal distribution against the timescale expressed in pulsar period.

We group the core-weak mode according to their time-scales, i.e. the period number of weak core mode, and obtained the quasi-regular patterns in the time-phase plots for related variations of profile components, see Figure~\ref{stablePatterns}. The core component get weakened to 14\% of the normal state, but the leading component get enhanced to 165\% of the normal state. \citet{tyw+2022} found that the weak core makes the peak ratio higher for the leading component with the trailing component, which lasts about 3 periods. That is clearly confirmed in our Figure \ref{coreNullProfile}. We find that the leading component is stronger than the normal state even at the end of core-weak mode, and then declines at least for 3 periods. At the same time, the trailing component is still weak, so that the peak ratio of the leading component over the trailing component get higher than the normal state. We find that the core-weak mode always trigger an intensity enhancement for some periods before and after the mode.

A core-weak profile can be fitted by 5 Gaussian components. According to the geometry method of PSR B0329+54 developed by  \citet{Gangadhara2004,Gangadhara2005}, the center of second cone, $(C_1 + C_5)/2-C_3$, should shifts toward an earlier longitude than that of the first cone does, i.e. $(C_2 + C_4)/2-C_3$. However, for the core-weak patterns, the  phase shift of the second cone is only $-1.03^\circ$, almost the same as that of the first cone, that is $-1.11^\circ$. 
The 5 components in core-weak patterns have been clearly detected at 325MHZ in the left-top subfigure of Figure 3 in \citet{MitraD2007_MNRAS} and at 1400MHZ in the left subfigure of Figure 2 in \citet{Brinkman2019}. That implies that these components are real over a broad-band. 

The most impressive characteristic of core-weak emission patterns is that the core-weak pattern starts and ends with the core component shifting and intensity enhancements. The similar phenomenon was found for PSR B0809+74 and PSR B0031-07 by \citet{Gajjar2014}, and PSR B0818-41 by \citet{Bhattacharyya2010} and PSR J1502$-$5653 by \citet{Li2012}. These pulsars null, and especially to an extreme nulling fraction of $>93\%$ for PSR J1502-5653. This implies that the core-weak phenomenon and the conventional nulling phenomenon have probably the same physical origin.

The explanations for the nulling phenomena are classified into two classes, the geometric models and the intrinsic models. In the geometric models,  \citet{Smits2005} thought that the sight line may intersect different part of pulsar magnetosphere for emission at a given frequency we are observing, so that the emission is missed and nulling occurs. The `core-null' or core-weak pattern  is a broad-band phenomenon, which is contradictory to such a model. \citet{Timokhin2010} pointed out that the nulling happens when the sight line gets out of the emission region which is shifted due to the magnetosphere shrinking, and the shape of magnetosphere is affected by the current flowing out along the open magnetic field lines \citep{Hibschman2001}.
In such a picture the magnetosphere shrinking can cause the weakening of emitted subpulses and subpulse shifting. However, this is hard to explain the phase difference of subpulses at the start and end of core-weak pattern, because the model of magnetosphere shrinking predicts a consecutive phase change of subpulses, like J1326-6700 \citep{WangN2007}. 
In the intrinsic models,  \citet{CR80} and  \cite{Gil2001} 
have developed a probable explanation for the nulling phenomenon based on the spark model \citep{RudermanMA1975_APJ}. \citet{Filippenko1982} thought that the nulling is caused by the emission mechanism switching, from the spark discharge to a steady discharge, which may not be able to create any instability conditions in the pulsar magnetosphere plasma, for example, via the two-stream instability. 
The physical processing to form the core-weak patterns of PSR B0329+54 is still mysterious. The core-weak emission patterns are probably triggered by a global current redistribution at a low height of the magnetosphere. A new model is needed to explain these quasi-regular patterns for the relatedly variation of the leading, core and trailing components.

\section*{Acknowledgements}
The authors are supported by the National Natural Science Foundation of
China (NSFC, Nos. 11988101, 11833009 and 12133004).

\section*{Data availability}
All data in this paper are available with kind request from authors. 





\bibliographystyle{mnras}
\bibliography{citation}


\appendix
\section{The Method to recognize the core-weak mode}
\label{appendix}

\begin{figure}
\centering
\includegraphics[width=0.75\linewidth]{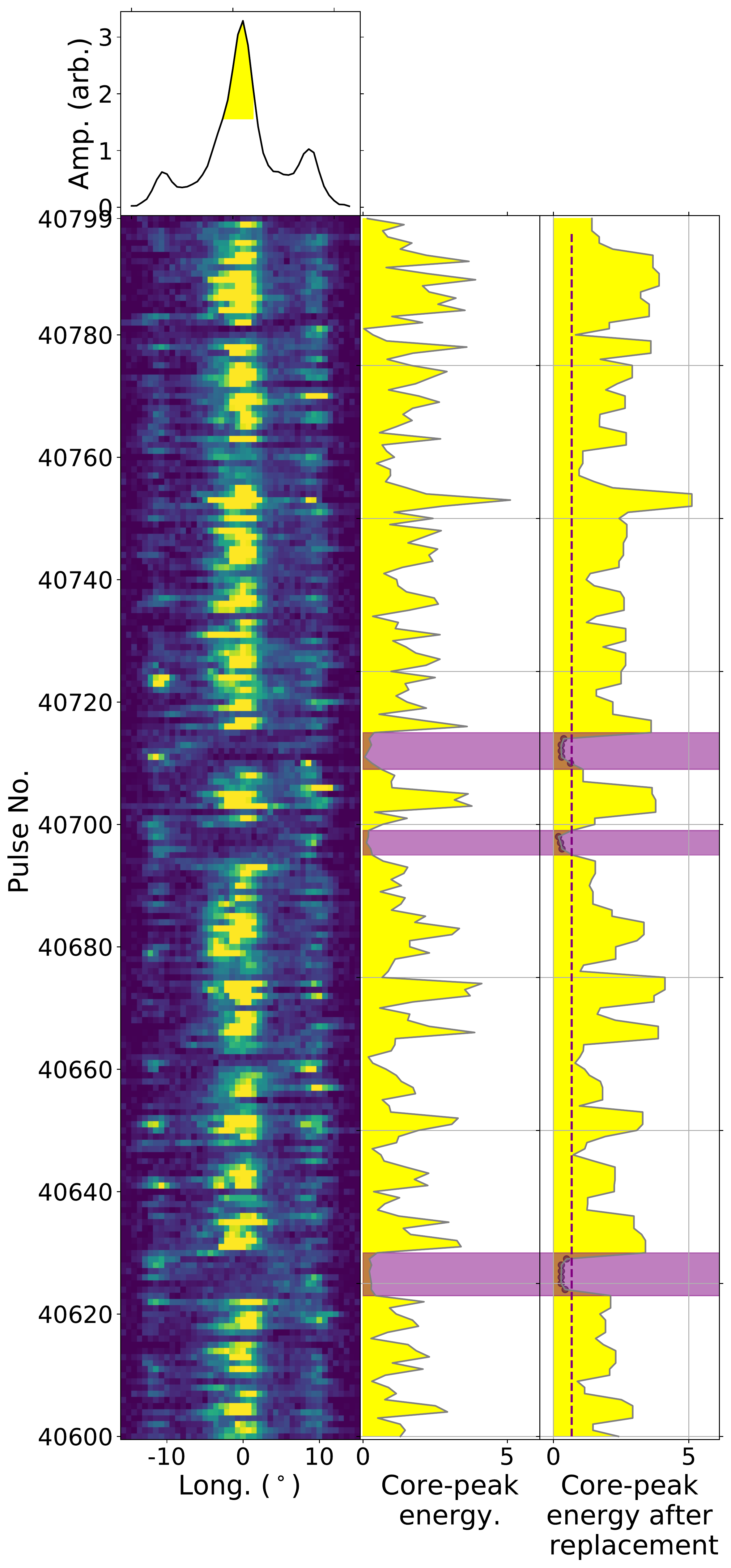}
\caption{A segment of data from the observations on 2017.11.08 on {\it the left panel}, as example, to show the variation of scaled core-peak energy for every period in {\it the middle panel}, and the 3-period energy-replacement values in {\it the right panel}. The core-peak energy for every period is integrated in the phase range of profiles above half peak (see the top sub-panel of pulse stacks. Three core-weak patterns with a longer timescale are recognized and marked, and the core-weak pulses in one or two periods are ignored. 
}
\label{coreNull}
\end{figure}

\begin{figure}
\centering
\includegraphics[width=0.8\linewidth]{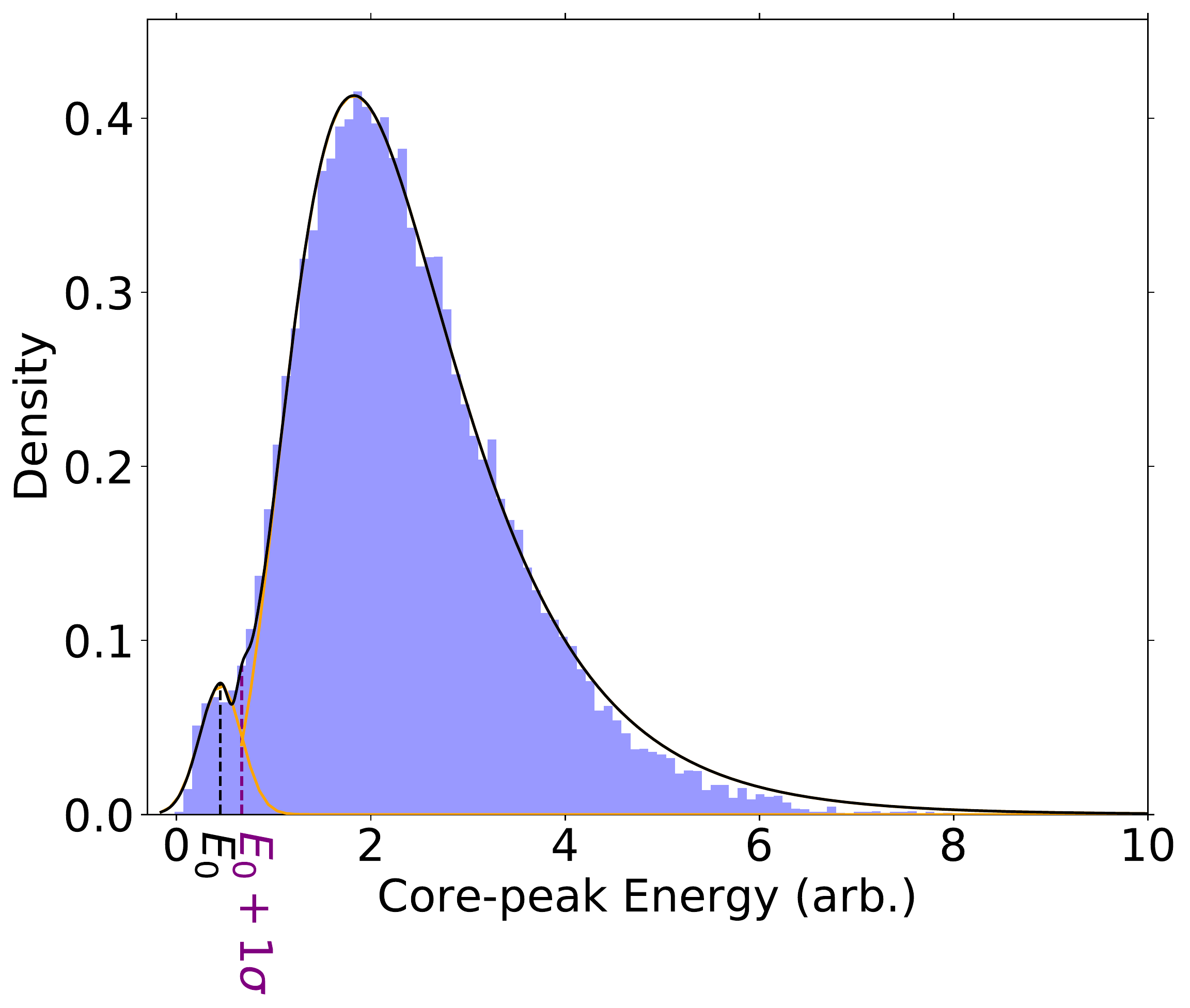}
\caption{The distribution of scaled ``Core-peak energy'' after the 3-period energy-replacement observed at 20171108. It can be fitted by a Gaussian function and a log-normal function. The Gaussian component is caused by the core-weak pulses, and the threshold of $< E_0+1\cdot\sigma$ can be used for recognition of the core-weak mode. We get the event-rates for core-weak mode about 4\% of periods by estimating the Gaussian component area.}
\label{coreHist}
\end{figure}

The core-weak mode have a characteristic of the core component gets weakened suddenly. However, the absolute amplitude changes of the core component depends many factors such as the interstellar scintillation. 

To filter out the core-weak emission mode with a simple criteria not affected by scintillation, the data must be normalized to remove the long-time modulation. The pulse energy of the core component in the phase range for the values above the half peak (see Figure~\ref{patternTemplate} and Figure~\ref{coreNull}), which we call it ``Core-peak energy'', are obtained for all individual pulses, and then the value at a given period is scaled by the root-mean-squares for $\pm100$ periods.  
Because the core-weak component keeps weak for at least 3 periods in the pattern, and the patterns with one or two periods of weak cores should not be considered here. We take two steps to find these core-weak mode. First, for a given period, the core-peak energy of this period $E_i$ is compared with those of the previous and the next periods, and is replaced by the largest among the three, i.e. 
\begin{equation}
     E_i = \max\{E_{i-1},E_{i},E_{i+1}\}.
\end{equation}
After such replacements, the low energy peak due to the random fluctuations would be diminished. Then the really low-energy core component for the pattern can be easily distinguished from the energy distributions of normal pulses, as showed in Figure~\ref{coreHist}. The core-peak energy distribution for pulses inside the core-weak patterns follows a Gaussian distribution, and that for the normal emission periods follows a log-normal distribution. We fitted the distribution with a Gaussian component and a log-normal component, and attribute these periods with a small core peak energy small than $E_0+1\sigma$ as a warranty criteria to pick up the core-weak mode. 

Taking the individual pulse data in Figure~\ref{coreNull} for example, we find that the scaled core-peak energy varies every period as shown in the middle panel, and that after the 3-period energy-replacement is shown in the right panel. The core-weak periods can be recognized as the red points lower than $E_0+1\sigma$. Some segments of data with a low Signal-Noise-Ratio has been discarded in the pattern recognition.  
This method can also be used to recognize nullings with a pulse series.


\bsp	
\label{lastpage}
\end{document}